\documentclass[a4paper,11pt]{article}

\usepackage[top=1.5in,right=1.0in,left=1.0in,bottom=1.0in]{geometry}

\usepackage{graphics,graphicx}
\usepackage{subfig}
\usepackage{float}
\usepackage{booktabs}
\usepackage[titletoc,title]{appendix}
\usepackage{longtable}
\usepackage{supertabular}
\usepackage{tabularx}
\usepackage{multirow}
\usepackage{array}
\usepackage{tabu}
\usepackage{mathtools}
\usepackage{hyperref}
\usepackage{cancel}

\newcommand{\otoprule}{\midrule[\heavyrulewidth]} 

\usepackage[normalem]{ulem}               
\newcommand\redout{\bgroup\markoverwith
	{\textcolor{red}{\rule[0.5ex]{2pt}{0.8pt}}}\ULon}
\usepackage{amsmath}
\usepackage{amsfonts}
\usepackage{amssymb}
\usepackage{siunitx} 

\usepackage{enumerate}
\usepackage{sectsty}
\usepackage[affil-it,auth-sc]{authblk}
\usepackage{color}
\usepackage[dvipsnames]{xcolor}

\sectionfont{\normalsize}
\subsectionfont{\normalsize}

\usepackage{tikz}
\usetikzlibrary{arrows,decorations,backgrounds,shapes, snakes}
\usetikzlibrary{positioning}
\usetikzlibrary{matrix} 
\usepackage{varwidth}
\usepackage[most]{tcolorbox}
\usetikzlibrary{shapes.geometric,arrows.meta,decorations.markings}
\usepackage{fancyhdr}            
\fancypagestyle{plain}{%
	\fancyhead{}                                     
	\fancyfoot[CE,CO]{\thepage}       
	\fancyhead[CE,CO]{\textcolor[rgb]{0.64,0.15,0.15}{Published in \emph{Journal of the Mechanics and Physics of Solids} (2020),  \textbf{134}, 103741 
			\\
			doi: \href{https://doi.org/10.1016/j.jmps.2019.103741}{10.1016/j.jmps.2019.103741}
	}}
	\setlength{\headheight}{14.5pt}}

\begin{document}

\newcommand{\singlespace}{\baselineskip=12pt\lineskiplimit=0pt\lineskip=0pt}
\def\ds{\displaystyle}

\newcommand{\beq}{\begin{equation}}
\newcommand{\eeq}{\end{equation}}
\newcommand{\lb}{\label}
\newcommand{\ph}{\phantom}
\newcommand{\beqar}{\begin{eqnarray}}
\newcommand{\eeqar}{\end{eqnarray}}
\newcommand{\barr}{\begin{array}}
\newcommand{\earr}{\end{array}}
\newcommand{\jump}{\parallel}
\newcommand{\Ehat}{\hat{E}}
\newcommand{\That}{\hat{\bf T}}
\newcommand{\Ahat}{\hat{A}}
\newcommand{\chat}{\hat{c}}
\newcommand{\shat}{\hat{s}}
\newcommand{\khat}{\hat{k}}
\newcommand{\muhat}{\hat{\mu}}
\newcommand{\mc}{M^{\scriptscriptstyle C}}
\newcommand{\mei}{M^{\scriptscriptstyle M,EI}}
\newcommand{\mec}{M^{\scriptscriptstyle M,EC}}
\newcommand{\hbeta}{{\hat{\beta}}}
\newcommand{\rec}[2]{\left( #1 #2 \ds{\frac{1}{#1}}\right)}
\newcommand{\rep}[2]{\left( {#1}^2 #2 \ds{\frac{1}{{#1}^2}}\right)}
\newcommand{\derp}[2]{\ds{\frac {\partial #1}{\partial #2}}}
\newcommand{\derpn}[3]{\ds{\frac {\partial^{#3}#1}{\partial #2^{#3}}}}
\newcommand{\dert}[2]{\ds{\frac {d #1}{d #2}}}
\newcommand{\dertn}[3]{\ds{\frac {d^{#3} #1}{d #2^{#3}}}}

\def\c{{\circ}}
\def\bob{{\, \underline{\overline{\otimes}} \,}}
\def\ob{{\, \underline{\otimes} \,}}
\def\scalp{\mbox{\boldmath$\, \cdot \, $}}
\def\gdp{\makebox{\raisebox{-.215ex}{$\Box$}\hspace{-.778em}$\times$}}
\def\daa{\makebox{\raisebox{-.050ex}{$-$}\hspace{-.550em}$: ~$}}
\def\mK{\mbox{${\mathcal{K}}$}}
\def\cK{\mbox{${\mathbb {K}}$}}

\def\Xint#1{\mathchoice
   {\XXint\displaystyle\textstyle{#1}}%
   {\XXint\textstyle\scriptstyle{#1}}%
   {\XXint\scriptstyle\scriptscriptstyle{#1}}%
   {\XXint\scriptscriptstyle\scriptscriptstyle{#1}}%
   \!\int}
\def\XXint#1#2#3{{\setbox0=\hbox{$#1{#2#3}{\int}$}
     \vcenter{\hbox{$#2#3$}}\kern-.5\wd0}}
\def\ddashint{\Xint=}
\def\fpint{\Xint=}
\def\dashint{\Xint-}
\def\cpvint{\Xint-}
\def\intl{\int\limits}
\def\cpvintl{\cpvint\limits}
\def\fpintl{\fpint\limits}
\def\ointl{\oint\limits}
\def\bA{{\bf A}}
\def\ba{{\bf a}}
\def\bB{{\bf B}}
\def\bb{{\bf b}}
\def\bc{{\bf c}}
\def\bC{{\bf C}}
\def\bd{{\bf d}}
\def\bD{{\bf D}}
\def\bE{{\bf E}}
\def\be{{\bf e}}
\def\bbf{{\bf f}}
\def\bF{{\bf F}}
\def\bG{{\bf G}}
\def\bg{{\bf g}}
\def\bi{{\bf i}}
\def\bH{{\bf H}}
\def\bK{{\bf K}}
\def\bL{{\bf L}}
\def\bM{{\bf M}}
\def\bN{{\bf N}}
\def\bn{{\bf n}}
\def\bo{{\bf o}}
\def\bm{{\bf m}}
\def\b0{{\bf 0}}
\def\bo{{\bf o}}
\def\bX{{\bf X}}
\def\bx{{\bf x}}
\def\bP{{\bf P}}
\def\bp{{\bf p}}
\def\bQ{{\bf Q}}
\def\bq{{\bf q}}
\def\bR{{\bf R}}
\def\bS{{\bf S}}
\def\bs{{\bf s}}
\def\bT{{\bf T}}
\def\bt{{\bf t}}
\def\bU{{\bf U}}
\def\bu{{\bf u}}
\def\bv{{\bf v}}
\def\bw{{\bf w}}
\def\bW{{\bf W}}
\def\by{{\bf y}}
\def\bz{{\bf z}}
\def\T{{\bf T}}
\def\Te{\textrm{T}}
\def\Id{{\bf I}}
\def\bxi{\mbox{\boldmath${\xi}$}}
\def\balpha{\mbox{\boldmath${\alpha}$}}
\def\bbeta{\mbox{\boldmath${\beta}$}}
\def\bepsilon{\mbox{\boldmath${\epsilon}$}}
\def\bvarepsilon{\mbox{\boldmath${\varepsilon}$}}
\def\bomega{\mbox{\boldmath${\omega}$}}
\def\bphi{\mbox{\boldmath${\phi}$}}
\def\bsigma{\mbox{\boldmath${\sigma}$}}
\def\bfeta{\mbox{\boldmath${\eta}$}}
\def\bDelta{\mbox{\boldmath${\Delta}$}}
\def\btau{\mbox{\boldmath $\tau$}}
\def\tr{{\rm tr}}
\def\dev{{\rm dev}}
\def\div{{\rm div}}
\def\Div{{\rm Div}}
\def\Grad{{\rm Grad}}
\def\grad{{\rm grad}}
\def\Lin{{\rm Lin}}
\def\Sym{{\rm Sym}}
\def\Skw{{\rm Skew}}
\def\abs{{\rm abs}}
\def\Re{{\rm Re}}
\def\Im{{\rm Im}}
\def\capB{\mbox{\boldmath${\mathsf B}$}}
\def\capC{\mbox{\boldmath${\mathsf C}$}}
\def\capD{\mbox{\boldmath${\mathsf D}$}}
\def\capE{\mbox{\boldmath${\mathsf E}$}}
\def\capG{\mbox{\boldmath${\mathsf G}$}}
\def\tcapG{\tilde{\capG}}
\def\capH{\mbox{\boldmath${\mathsf H}$}}
\def\capK{\mbox{\boldmath${\mathsf K}$}}
\def\capL{\mbox{\boldmath${\mathsf L}$}}
\def\capM{\mbox{\boldmath${\mathsf M}$}}
\def\capR{\mbox{\boldmath${\mathsf R}$}}
\def\capW{\mbox{\boldmath${\mathsf W}$}}

\def\i{\mbox{${\mathrm i}$}}
\def\mC{\mbox{\boldmath${\mathcal C}$}}
\def\mB{\mbox{${\mathcal B}$}}
\def\mE{\mbox{${\mathcal{E}}$}}
\def\mL{\mbox{${\mathcal{L}}$}}
\def\mK{\mbox{${\mathcal{K}}$}}
\def\mV{\mbox{${\mathcal{V}}$}}
\def\C{\mbox{\boldmath${\mathcal C}$}}
\def\E{\mbox{\boldmath${\mathcal E}$}}

\def\AAM{{\it Advances in Applied Mechanics }}
\def\ACME{{\it Arch. Comput. Meth. Engng.}}
\def\ARMA{{\it Arch. Rat. Mech. Analysis}}
\def\AMR{{\it Appl. Mech. Rev.}}
\def\ASCEEM{{\it ASCE J. Eng. Mech.}}
\def\ACTA{{\it Acta Mater.}}
\def\CMAME {{\it Comput. Meth. Appl. Mech. Engrg.}}
\def\CRAS{{\it C. R. Acad. Sci. Paris}}
\def\CRM{{\it Comptes Rendus M\'ecanique}}
\def\EFM{{\it Eng. Fracture Mechanics}}
\def\EJMA{{\it Eur.~J.~Mechanics-A/Solids}}
\def\IJES{{\it Int. J. Eng. Sci.}}
\def\IJF{{\it Int. J. Fracture}}
\def\IJMS{{\it Int. J. Mech. Sci.}}
\def\IJNAMG{{\it Int. J. Numer. Anal. Meth. Geomech.}}
\def\IJP{{\it Int. J. Plasticity}}
\def\IJSS{{\it Int. J. Solids Structures}}
\def\IngA{{\it Ing. Archiv}}
\def\JAM{{\it J. Appl. Mech.}}
\def\JAP{{\it J. Appl. Phys.}}
\def\JAE{{\it J. Aerospace Eng.}}
\def\JE{{\it J. Elasticity}}
\def\JM{{\it J. de M\'ecanique}}
\def\JMPS{{\it J. Mech. Phys. Solids}}
\def\JSV{{\it J. Sound and Vibration}}
\def\MACRO{{\it Macromolecules}}
\def\MMT{{\it Mech. Mach. Th.}}
\def\MOM{{\it Mech. Materials}}
\def\MMS{{\it Math. Mech. Solids}}
\def\MMT{{\it Metall. Mater. Trans. A}}
\def\MPCPS{{\it Math. Proc. Camb. Phil. Soc.}}
\def\MSE{{\it Mater. Sci. Eng.}}
\def\NATURE{{\it Nature}}
\def\NATUREM{{\it Nature Mater.}}
\def\PHIL{{\it Phil. Trans. R. Soc.}}
\def\PMPS{{\it Proc. Math. Phys. Soc.}}
\def\PNAS{{\it Proc. Nat. Acad. Sci.}}
\def\PRE{{\it Phys. Rev. E}}
\def\PRL{{\it Phys. Rev. Letters}}
\def\PRSL{{\it Proc. R. Soc.}}
\def\ROCK{{\it Rock Mech. and Rock Eng.}}
\def\QAM{{\it Quart. Appl. Math.}}
\def\QJMAM{{\it Quart. J. Mech. Appl. Math.}}
\def\SCIENCE{{\it Science}}
\def\SCRMAT{{\it Scripta Mater.}}
\def\SM{{\it Scripta Metall.}}
\def\ZAMM{{\it Z. Angew. Math. Mech.}}
\def\ZAMP{{\it Z. Angew. Math. Phys.}}
\def\ZVDI{{\it Z. Verein. Deut. Ing.}}

\renewcommand\Affilfont{\itshape\small}

\title{
Structures loaded with a force acting along a fixed straight line, \\ or the \lq \lq Reut's column problem''
}

\author[1]{Davide Bigoni\footnote{Corresponding author:\,e-mail:\, davide.bigoni@ing.unitn.it; phone:\,+39\,0461\,282507.}}
\author[1]{Diego Misseroni}
\affil[1]{Department of Civil, Environmental and Mechanical Engineering, University of Trento\\via Mesiano 77, Trento, Italy}

\date{}
\maketitle

\begin{abstract}
An elastic double pendulum subject to a force acting along a fixed straight line, the so-called \lq \lq Reut's column problem'', is a structure 
exhibiting flutter and divergence instability, which was never realized in practice and thus 
debated whether to represent reality or mere speculation. 
It is shown, both theoretically and experimentally, how to obtain the Reut's loading by exploiting the contact with friction of a rigid blade 
against a freely-rotating cylindrical constraint, which moves axially at constant speed,  
an action recalling that of a bow's hair on a violin string. With this experimental set-up, flutter and divergence instabilities, as well as the detrimental effect of viscosity on critical loads, are documented 
indisputably, thus bringing an end to a long debate. This result opens a new research area, with perspective applications to mechanical actuators, high-precision cutting tools, or energy harvesting devices.

\end{abstract}

\noindent Keywords: Coulomb friction; flutter instability; divergence instability; follower load; Hopf bifurcation;

\section{Introduction} 

\begin{figure}[h]
  \begin{center}
\includegraphics[width=0.8\textwidth]{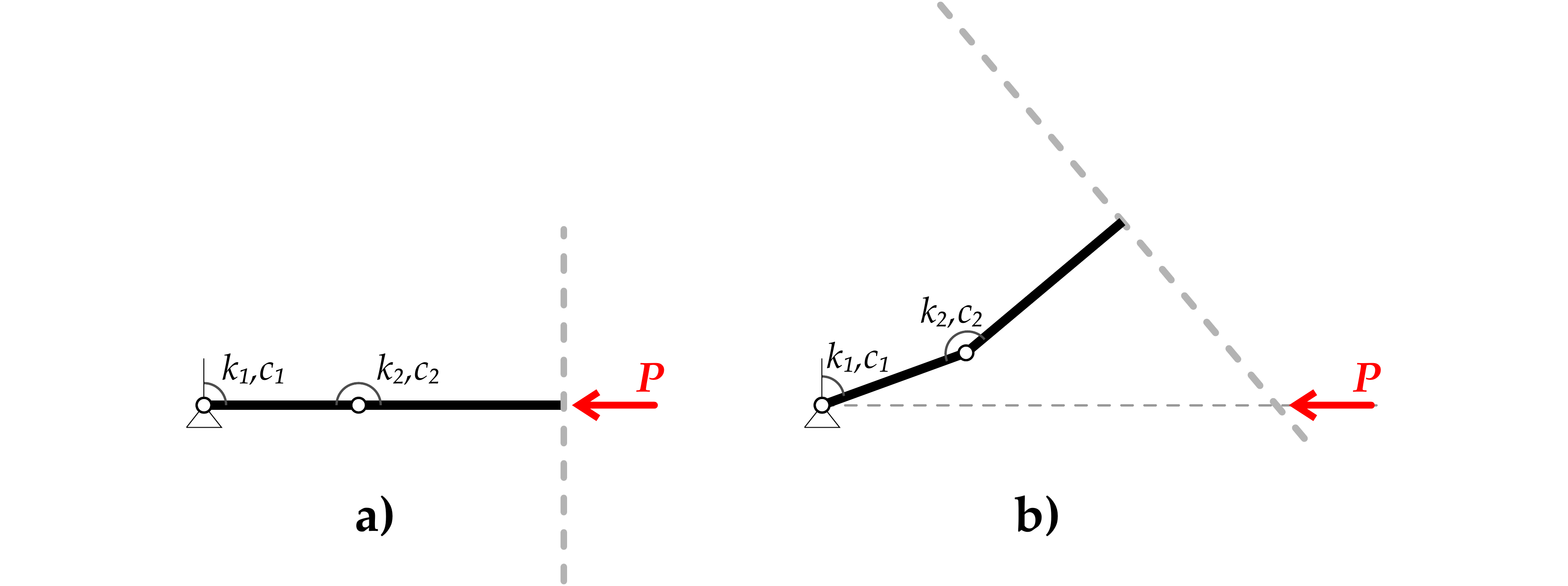}
\caption{\footnotesize An elastic double pendulum loaded by a force constrained to remain coaxial to a fixed straight line, but allowed to freely slide against a rigid blade (sketched dashed/grey in the figure), the so-called \lq Reut's column'. This load 
triggers flutter and divergence instability in the structure which occur at the same values of critical load as in the Ziegler double pendulum.
}
\label{reutotto}
 \end{center}
\end{figure}

A force constrained to act on an elastic structure while remaining on a given straight line is an example of non-conservative load, which was introduced through the so-called \lq \lq Reut's column problem'' \cite{reut}, sketched in Fig. \ref{reutotto}. 
The column is modeled as an elastic double pendulum with two viscoelastic hinges of stiffnesses $k_1$ and $k_2$ and dampings $c_1$ and $c_2$, as in the case investigated by Ziegler \cite{ziegler_0}. 
The force $P$ is left free to slide along a rigid blade (the dashed/grey element shown in the figure), but has to lie always on the same horizontal line. 
It can be shown \cite{bolotin, herrmannb} that the Reut's column is subject to flutter and divergence instabilities and that its critical loads are the same as those
calculated for the Ziegler double pendulum subject to a tangential follower force \cite{ziegler_0}. It is perhaps less known that the Reut's column also exhibits the Ziegler paradox, which corresponds to a discontinuity in the critical load 
occurring in the limit of null viscosity \cite{ziegler_0, bolotin, bigoni-cism, bottema, kirillov_1, kirillov_libro, kirillov_2, kirillov_3}.
%

The difficulty with the Reut's load lies in its realization and on this point Bolotin \cite{bolotin} writes: \textit{ \lq \lq It is clear that something similar to such a force 
could be produced by the pressure from a jet of absolutely inelastic particles''}. 
However, inelastic particles do not exist in practice and, if these remain attached after impact, the mass of the structure changes, so that the proposal 
by Bolotin is simply unrealistic. 
Hermann et al. \cite{herrmannb} 
and Sugiyama \cite{sugiyama_mi, sugiyamasolo} used an air jet impinging the structure to produce the load, a set-up much different from that evisaged by Bolotin. Flutter instability was found 
by the former research team \cite{herrmannb}, while Sugiyama was initially unable to find it \cite{sugiyama_mi}, but succeeded later \cite{sugiyamasolo}. 
In any case, the air jet produces with the structure a complex aeroelastic interaction, leading to a load differing from that idealized by Reut.
Therefore, the force postulated by Reut remained until now (for 80 years!) a boundary condition with nothing but a mathematical meaning, never observed in the real world (see the vivid description
of the state-of-the-art given in \cite{elishakoff}).
A similar difficulty arose in the realization of the tangentially follower force introduced by Ziegler, which, considered \lq unrealistic' by Koiter \cite{koiter, elishakoff}, was only recently realized by 
Bigoni and Noselli \cite{bigoninoselli} (see also \cite{bigkir1, bigkir2}) through the introduction of a special constraint, namely, a freely rotating
wheel sliding with friction against a moving rigid plate. 

\begin{figure}[h]
  \begin{center}
\includegraphics[width=0.8\textwidth]{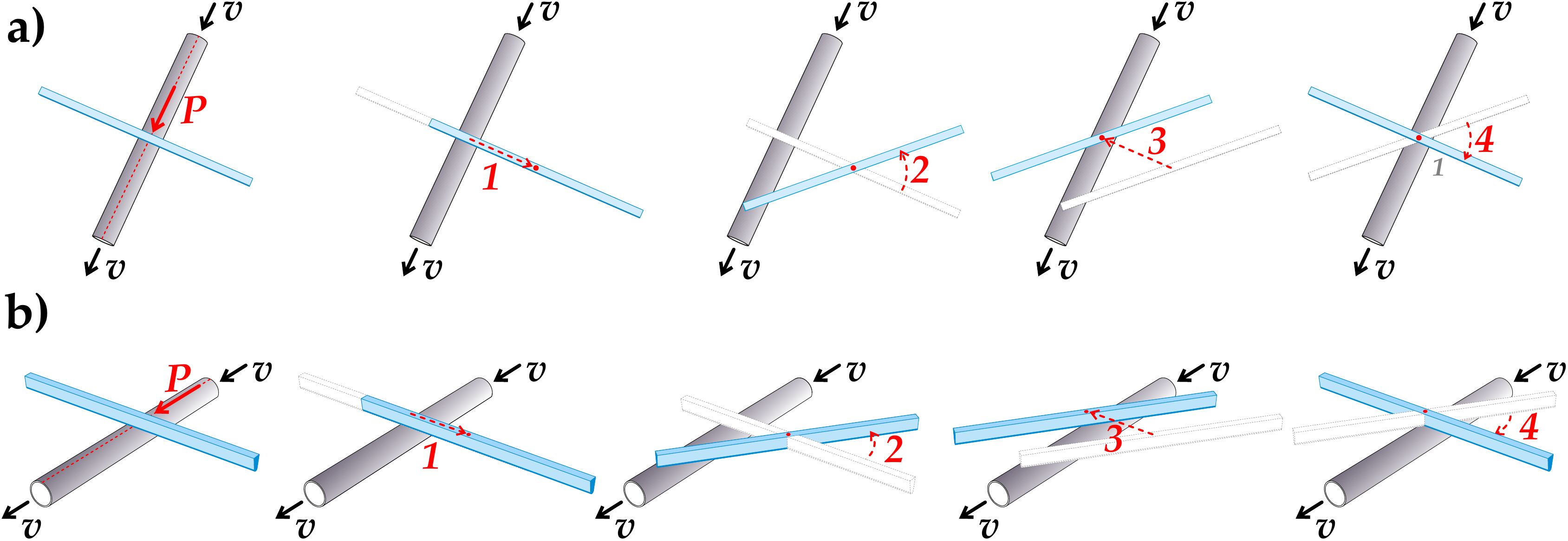}
\caption{\footnotesize Two equivalent views (a) and (b) show the interaction between a rigid and rough cylindrical surface sliding with constant speed $v$ against a rigid blade, which produces a frictional force $P$ remaining parallel to the axis of the cylinder (on the left). This force provides the load postulated by Reut and is able to produce positive work in a closed cycle, as illustrated by the sequence of figures (from left to right): initially the blade slides orthogonally to the cylinder without producing any work (movement 1); then it rotates anticlockwise about its centre  absorbing from $P$ a positive work (because the force is displaced parallel to itself, movement 2); finally, translation orthogonal to the cylinder (movement 3) and rotation around the contact point (movement 4) produces no further work and completes a closed loop in which work has been transmitted to the blade from the sliding with friction of the cylinder.}
\label{lo_chiamavano_reut}
 \end{center}
\end{figure}

In this article it is shown that the Reut's load can be obtained through the sliding, with constant Coulomb friction, of a cylindrical 
surface free of rotating about its axis against a rigid blade, as shown from two equivalent views (but taken with a different perspective) in Fig. \ref{lo_chiamavano_reut}. The cylindrical constraint is idealized with negligible mass so that it can transmit to the blade only an axial force. In fact, the free-rotation condition does not allow the generation of tangential actions orthogonal to the cylinder's axis. 
%
The action of the cylindrical constraint on the blade is in a sense similar to the action of the hair of a bow on violin strings. 

As shown in Fig. \ref{lo_chiamavano_reut}, the introduced frictional constraint can produce a positive work in a closed cycle and transmit it to the elastic structure, so that the steady energy input 
provided by the cylinder transforms the structure in a self-oscillating system when flutter instability or divergence instability occur. The way of realizing the 
Reut's load (\lq cylinder-blade contact') is dual to how the Ziegler's load (\lq wheel-plane contact') has been obtained by Bigoni and Noselli \cite{bigoninoselli}, so that it is 
suggested that these two possibilities of generating nonconservative loads are in a sense unique and complementary. 
 
It may be highlighted that the Reut's force cannot be produced through contact with a frictionless guided weight subject to gravity, as the frictionless condition implies that the 
force transmitted to the rigid blade be orthogonal to it. 
This loading condition will also be considered, in this article in addition to 
dead loading, and follower loading (in the Ziegler sense), three different loads included to facilitate comparisons.

\begin{figure}[h]
	\begin{center}
		\includegraphics[width=0.8\textwidth]{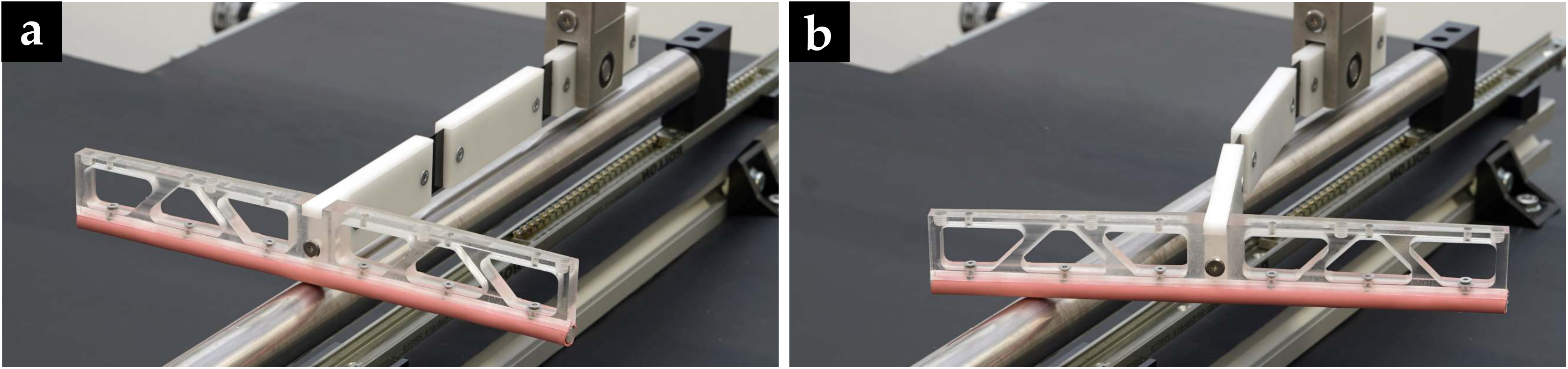}
		\caption{\footnotesize Schematics of the experimental set-up for applying a force acting on a straight line to an elastic double pendulum: a freely-rotating cylindrical constraint is sliding with friction against a rigid element of the structure. The structure is in the initial (a) and in a deformed (b) configuration.}
		\label{fig_schema0}
	\end{center}
\end{figure}

\begin{figure}[h]
	\begin{center}
		\includegraphics[width=0.8\textwidth]{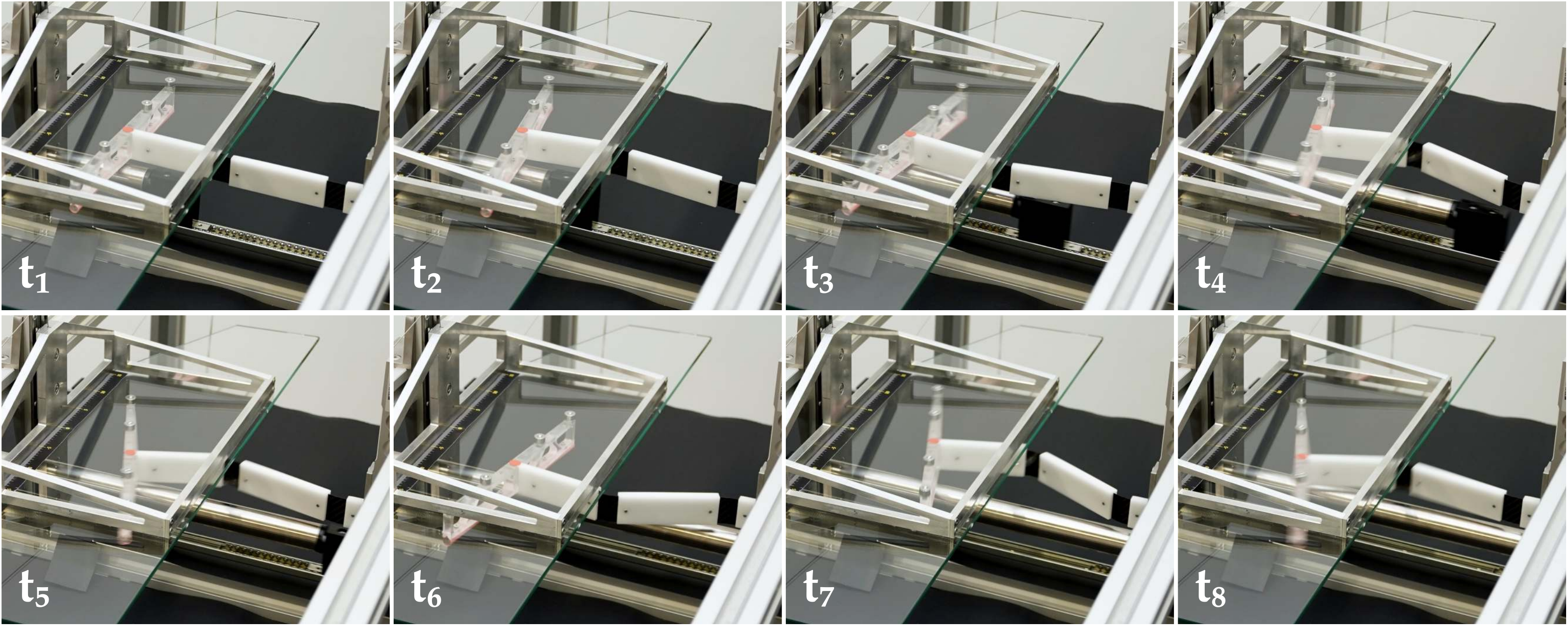}
		\caption{\footnotesize A sequence of photos during an experiment documenting flutter instability in the Reut's column.}
		\label{sequenza}
	\end{center}
\end{figure}

The ideal scheme illustrated in Fig. \ref{lo_chiamavano_reut} to produce the Reut's load will be shown to be implementable in reality (the experimental set-up is shown in 
Fig. \ref{fig_schema0}). 
Experimental results documenting flutter and divergence instabilities in the Reut's column will be presented (a sequence of photos taken during an experiment is anticipated in Fig. \ref{sequenza}, see also the electronic supplementary material and
also http://ssmg.ing.unitn.it/), showing how the contact with the rough cylindrical surface may induce flutter.

Our experiments are in excellent agreement with the theoretical results, providing measured values of the flutter and divergence instability thresholds lying between the two values corresponding to the viscoelastic and purely elastic schemes. These experiments, therefore, provide another experimental proof 
of the destabilization by damping, so supporting the concept of dissipative instability \cite{kre} and related Ziegler paradox. 

The introduced mechanical set-up shows for the first time after the 80-years-old Reut's paper, that a load acting on a fixed line can be made a
real and usable condition, thus opening the way to new and unexpected applications, for instance in mechanical actuation or energy harvesting \cite{xia}, but also in models for locomotion \cite{Bayly}, biomechanics \cite{canio}, and fluid-structure interaction \cite{maha}, research arenas where flutter instability may play an important role.

\section{Elastic double pendulum subject to four different loads of conservative and nonconservative nature}

\begin{figure}[h]
  \begin{center}
\includegraphics[width=0.8\textwidth]{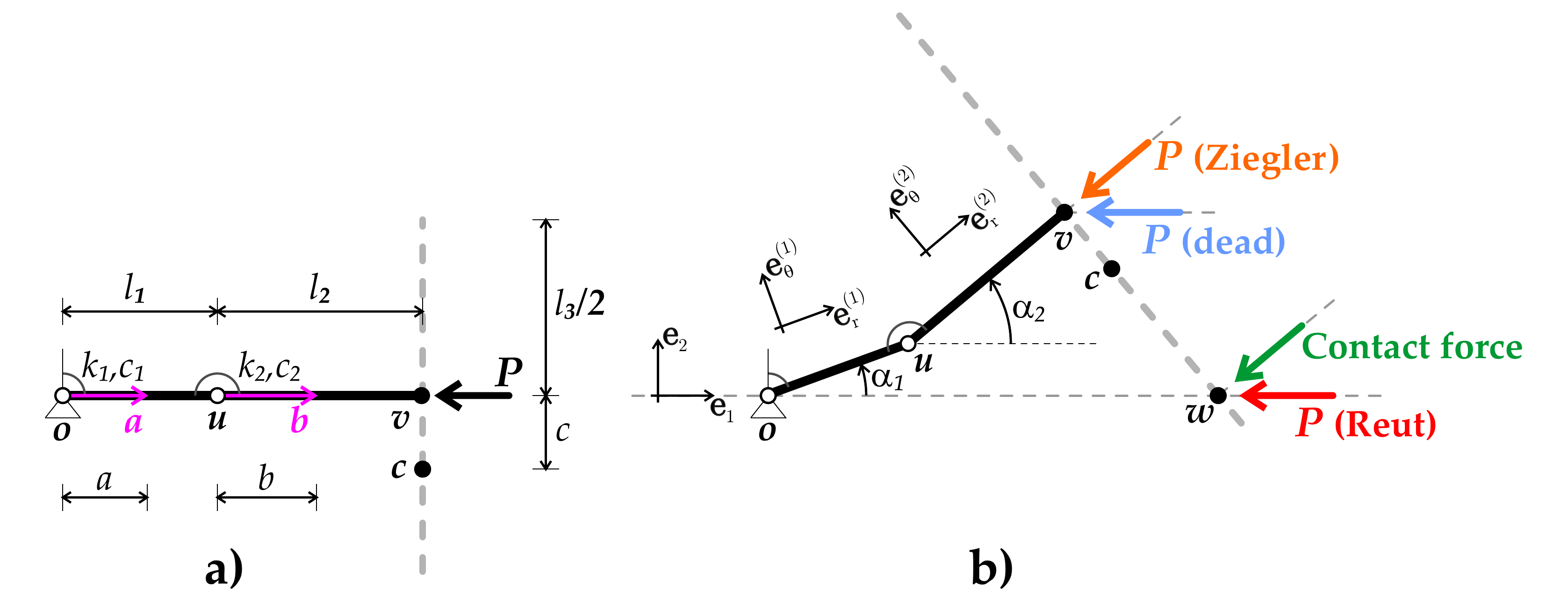}
\caption{\footnotesize A double pendulum (in the reference configuration, left,  and in a deformed configuration, right) with heavy rigid bars and visco-elastic hinges (located at $\bo$ and $\bu$) and 
subject to four types of horizontal forces: (i.) a horizontal \lq dead' load $P$ applied at $\bv$;
(ii.) the \lq Reut' non-conservative load $P$, remaining parallel to the axis $\be_1$; (iii.) a \lq contact' force of horizontal fixed component $P$ is applied parallel to the axis $\be_1$; 
(iv.) the \lq Ziegler' non-conservative load $P$ applied at the point $\bv$ and remaining parallel to the rod $\bv-\bu$. Note the two degrees of freedom $\alpha_1$ and $\alpha_2$ and that the rigid \lq blade' orthogonal to the bar $\bv-\bu$ is only needed to transmit the two loads (ii.) and (iii.). 
}
\label{reutinino}
 \end{center}
\end{figure}

A double pendulum is considered, namely, the two-degree-of-freedom rigid and heavy (the rods have unit mass densities $\rho_1$, $\rho_2$, and $\rho_3$) rods system shown in Fig. \ref{reutinino}. 
The system is made elastic by 
two rotational springs of stiffnesses $k_1$ and $k_2$, which may also display a viscous behaviour 
with coefficients $c_1$ and $c_2$, and will be subject to four different loads, all initially 
applied at the junction point $\bv$. Note that the \lq blade', in other words the bar orthogonal (and rigidly attached) to the element $\bv-\bu$, is only needed to transmit two of the considered loads to the system. 

\paragraph{The kinematic description of the double pendulum} does not need considering the (visco-elastic) constitutive law of the hinges and the applied loads. 
%
The generic configuration 
of the system remains determined by the two Lagrangean angles $\alpha_1$ and $\alpha_2$, so that 
denoting 
the time derivative by a superimposed dot, the kinematics of the double pendulum can be obtained from the knowledge of the position, velocity and acceleration of points $\ba$ (belonging to the 
bar $\bu-\bo$), $\bb$ (belonging to the bar $\bv-\bu$), and $\bc$ (belonging to the bar orthogonal to the element 
$\bv-\bu$). Point $\bc$ moves in agreement to the following rules 
\beq
\lb{strasoccia}
\begin{array}{lll}
\bc = l_1 \be_r^{(1)} + l_2 \be_r^{(2)} - c \be_\theta^{(2)} + \bo, \\ [3 mm]
\dot{\bc}= \ds l_1 \dot{\alpha}_1\be_\theta^{(1)} 
+ l_2 \dot{\alpha}_2\be_\theta^{(2)} 
+ c \dot{\alpha}_2 \be_r^{(2)}, \\ [3 mm]
\ddot{\bc}= 
l_1 \ddot{\alpha_1}\be_\theta^{(1)} - l_1 \dot{\alpha}_1^2\be_r^{(1)} 
+ l_2 \ddot{\alpha}_2\be_\theta^{(2)} - l_2 \dot{\alpha}_2^2\be_r^{(2)} 
+ c \ddot{\alpha}_2 \be_r^{(2)} + c \dot{\alpha}_2^2 \be_\theta^{(2)},
\end{array}
\eeq
where $\be_r^{(j)}$, $\be_\theta^{(j)}$ (with $j=1,2$) are unit vectors respectively 
parallel and orthogonal to the bar $j$, Fig. \ref{reutinino}. Placement, velocity, and acceleration of points $\ba$ and $\bb$ are specified in Appendix \ref{appendixa}.

Note that for the Reut's and the contact cases, the load is applied at point $\bw$, so that 
the distance  $c$ between points $\bc$ and $\bv$ on the blade, when $\bc$ is 
momentarily superimposed with $\bw$, is 
\beq
\lb{oca}
c= \frac{l_1\sin\alpha_1+l_2\sin\alpha_2}{\cos\alpha_2},
\eeq
so that the velocity of the point $\bw$ of the rigid arm momentarily intersecting the $x_1$--axis is obtained as
\beq
\lb{catzo}
\dot{\bw} = l_1\sin\alpha_1 \left(-\dot{\alpha}_1+ \dot{\alpha}_2\right) \be_1 + l_1 \left(\dot{\alpha}_1 \cos\alpha_1 + \dot{\alpha}_2 \frac{\sin\alpha_1 \sin\alpha_2 +l_2/l_1}{\cos\alpha_2} \right)
\be_2 , 
\eeq
which, in general, is {\it not} parallel to $\be_1$. 

\paragraph{The constitutive equations for the hinges} connecting the rigid bars are assumed visco-elastic and defined by the stiffnesses $k_j$ and $c_j$ ($j=1,2$), so that   
the moments transmitted by the rotational springs to the rods are $k_1 \alpha_1+c_1 \dot{\alpha}_1$ and 
$k_2 (\alpha_2-\alpha_1) +c_2 (\dot{\alpha}_2-\dot{\alpha}_1)$.

\paragraph{The principle of virtual powers for the double pendulum} writes as
\beq
\lb{virtuale}
\begin{array}{ll}
\ds \underbrace{\bP \scalp \delta \bw}_{external~power} & = 
\underbrace{(k_1 \alpha_1 + c_1 \dot{\alpha}_1)\delta \alpha_1 + \left[k_2 (\alpha_2-\alpha_1)
+ c_2 (\dot{\alpha}_2-\dot{\alpha}_1)\right]
(\delta \alpha_2-\delta \alpha_1)}_{internal~power}
 \\ [5 mm]
~ & \ds + \underbrace{\rho_1\int_0^{l_1} \left(\ddot{\ba}\scalp\delta\ba \right) da
+ \rho_2\int_0^{l_2} \left(\ddot{\bb}\scalp\delta\bb \right) db
+ \rho_3\int_{-l_3/2}^{l_3/2} \left(\ddot{\bc}\scalp\delta\bc \right) dc}_{inertia} ,
\end{array}
\eeq
where 
the virtual velocities $\delta \ba$, $\delta \bb$, $\delta \bc$, 
have the same expressions (\ref{strasoccia2})$_2$, (\ref{strasoccia2})$_5$, and
(\ref{strasoccia})$_2$ with the \lq $\dot{~}$' replaced by \lq $\delta$'.

The virtual power of an external load applied at a point $\bc$ (singled out by the distance $c$ between $\bv$ and $\bc$) can be written as
\beq
\lb{kariko}
\bP\scalp\delta\bw = l_1 \bP\scalp\be_\theta^{(1)} \, \delta\alpha_1
+\left( l_2 \bP\scalp\be_\theta^{(2)} + c \bP\scalp\be_r^{(2)}\right)\delta\alpha_2. 
\eeq

\paragraph{Four loads are applied to the double pendulum,} two conservative 
(a \lq dead' and a \lq frictionless contact') and two nonconservative loads (the \lq Ziegler' and the \lq Reut'). Consideration of these loads, separately acting on the structure, is particularly instructive.  

For all cases of loading, 
the  
force $\bP$ 
transmitted to the structure 
(referred to 
the unit vectors $\be_1$ and $\be_2$ singling out the horizontal and vertical direction respectively, Fig. \ref{reutinino}) are
\beq
\lb{caricazzi}
\bP = -P 
\left\{
\begin{array}{lll}
~ \be_1   & ~~~~ \mbox{ Dead loading} \\ 
\ds \left(\be_1 + \tan \alpha_2 \, \be_2 \right)\cos\alpha_2  & ~~~~ \mbox{ Ziegler} \\ 
\ds ~\be_1 + \tan \alpha_2 \, \be_2 &~~~~\mbox{ Frictionless contact} \\ 
~ \be_1   & ~~~~ \mbox{ Reut loading} \\ 
\end{array}
\right.
\eeq
where 
$P$ is the constant modulus of the force $\bP$, except in the case of the frictionless contact, where $P$ is the component of $\bP$ along $\be_1$, or the modulus of $\bP$ when $\alpha_2=0$.

It is clear from (\ref{caricazzi}) that the applied forces for the dead and Reut loads are identical and that the forces for the Ziegler and frictionless contact coincide only in a linearization written for 
small deviations from the trivial equilibrium configuration 
\beq
\lb{linka}
\bP = -P 
\left\{
\begin{array}{lll}
\be_1   & ~~~~ \mbox{ Reut or dead loading,} \\ 
\ds \be_1 + \alpha_2 \, \be_2 & ~~~~ \mbox{ Ziegler or frictionless contact.} 
\end{array}
\right.
\eeq

It can be concluded from expressions (\ref{caricazzi}) and (\ref{linka}) that  it is not simply the form of the force which decides 
if a mechanical system
does or does not admit a potential structure.

The treatments of dead loading, loading through frictionless contact, and the Ziegler load are deferred to Appendix \ref{appendixa}, while the Reut load is analyzed below.

\paragraph{The Reut load} is nonconservative and is characterized as follows:  the force $\bP$ is (i.) free of moving  
along the bar orthogonal to $\bv-\bu$, but has to (ii.) belong and (iii.) remain  parallel to the 
$\be_1$-axis. 
Such a force does not follow the structure and has the same direction  as the dead load. 
As for the frictionless contact load, the load is always applied at the point $\bw$, so that the two points $\bc$ (belonging to the double pendulum) and $\bw$ (belonging to the reference 
$\be_1$-axis) are 
momentarily superimposed when Eq. (\ref{oca}) holds. 

The power of external load can be obtained from Eq. (\ref{kariko}), by setting for $c$ the value (\ref{oca}) and thus obtaining
\beq
\lb{straza}
\bP \scalp \delta\bw = P l_1 \sin\alpha_1
\left( \delta \alpha_1- \delta\alpha_2\right) ,
\eeq
so that the load does not admit a potential because
$
\partial \sin\alpha_1/\partial \alpha_2\neq
\partial (-\sin\alpha_1)/\partial \alpha_1.
$
The power of the Reut load (\ref{straza}) can be linearized near the trivial equilibrium position as
\beq
\lb{kariko-reut}
\bP \scalp \delta\bw = P l_1 \alpha_1 \left( \delta \alpha_1- \delta\alpha_2\right).
\eeq

\paragraph{The equations of motion for a double pendulum}
can be obtained from the virtual power principle (\ref{virtuale}), by invoking the arbitrariness of $\delta \alpha_1$ and $\delta \alpha_2$, which 
yields the two equations
\beq
\lb{nonlinearazzo}
\begin{array}{ll}
\ds \left(\rho_1\frac{l_1^3}{3} +  \rho_3 l_1^2l_3 +\rho_2l_1^2l_2\right)
\ddot{\alpha}_1 
+ \left(\rho_3l_1l_2l_3+\rho_2\frac{l_1l_2^2}{2}\right) \ddot{\alpha}_2 \cos{(\alpha_1-\alpha_2)} \\ [5 mm]
\ds + \left(\rho_3l_1l_2l_3+\rho_2\frac{l_1l_2^2}{2}\right) \dot{\alpha}_2^2 \sin{(\alpha_1-\alpha_2)} 
+ k_1 \alpha_1 + k_2 (\alpha_1-\alpha_2) 
+ c_1 \dot{\alpha}_1 + c_2 (\dot{\alpha}_1-\dot{\alpha}_2) 
- l_1 \bP\scalp\be_\theta^{(1)} = 0, \\ [5 mm]
\ds \left(\rho_3l_1l_2l_3+\rho_2\frac{l_1l_2^2}{2}\right)  \ddot{\alpha}_1 \cos{(\alpha_1-\alpha_2)} 
+ \left[\rho_3\left(l_2^2l_3+\frac{l_3^3}{12}\right)+\rho_2\frac{l_2^3}{3}\right] \ddot{\alpha}_2  \\ [5 mm]
\ds ~~~~~~~~
-l_1(\rho_3l_2l_3+\rho_2\frac{l_2^2}{2}) \dot{\alpha}_1^2 \sin{(\alpha_1-\alpha_2)} 
 - k_2 (\alpha_1-\alpha_2) - c_2 (\dot{\alpha}_1-\dot{\alpha}_2)- \ds l_2 \bP\scalp\be_\theta^{(2)} - c \bP\scalp\be_r^{(2)}= 0 ,
\end{array}
\eeq
governing the nonlinear dynamics of the double pendulum.

The differential equations (\ref{nonlinearazzo}), linearized near the trivial (equilibrium) 
configuration $\alpha_1=\alpha_2=0$, become
\beq
\lb{linearizzatissime}
\begin{array}{ll}
\ds \left[\rho_1\frac{l_1^3}{3} +  \rho_3 l_1^2l_3 +\rho_2l_1^2l_2\right]
\ddot{\alpha}_1 
+ \left(\rho_3l_1l_2l_3+\rho_2\frac{l_1l_2^2}{2}\right) \ddot{\alpha}_2  
+ k_1 \alpha_1 + k_2 (\alpha_1-\alpha_2) \\ [5 mm]
~~~~~~~~~~~~~~ + c_1 \dot{\alpha}_1 + c_2 (\dot{\alpha}_1-\dot{\alpha}_2) 
- l_1 P_{w1} = 0, \\ [5 mm]
\ds \left(\rho_3l_1 l_2l_3+\rho_2\frac{l_1 l_2^2}{2}\right) \ddot{\alpha}_1 
+ \left[\rho_3\left(l_2^2l_3+\frac{l_3^3}{12}\right)+\rho_2\frac{l_2^3}{3}\right] \ddot{\alpha}_2  
- k_2 (\alpha_1-\alpha_2) - c_2 (\dot{\alpha}_1-\dot{\alpha}_2)  -l_2 P_{w2} =0 ,
\end{array}
\eeq
 
The terms defining the external power in equation (\ref{nonlinearazzo}) and in its linearized version (\ref{linearizzatissime}) are reported in Table  \ref{externalpower}
with reference to all four considered loads.

\begin{table}[H]
	\footnotesize
	\renewcommand{\tablename}{\footnotesize{Tab.}}
	\centering
	\begin{tabu}  to 1\linewidth {X[1.1c]| *{1}{X[2$c$]} *{1}{X[1.4$c$]}| *{1}{X[2$c$]} *{1}{X[1.4$c$]}}
		\otoprule
		  & \multicolumn{4}{c}{\textbf{External Power}}  \\[1mm]
		  &  \text{Non linear} & \text{Linearized} & \text{Non linear} & \text{Linearized}  \\[1mm]
		  \textbf{Force type} & \bP\scalp\be_\theta^{(1)} & P_{w1}  & \ds  \bP\scalp\be_\theta^{(2)} + c/l_2 \bP\scalp\be_r^{(2)}& P_{w2 }  \\[2mm]
		\hline
		Dead  &  P\sin\alpha_1 & P\alpha_1    & P\sin\alpha_2 & P\alpha_2 \\[1mm]
		Ziegler  &  P\sin(\alpha_1-\alpha_2) &P(\alpha_1-\alpha_2)   & 0  & 0 \\ [1mm]
		Contact   & P\left(\sin\alpha_1-\cos\alpha_1\tan\alpha_2\right) &  P\left(\alpha_1-\alpha_2\right)   & \ds -P\frac{l_1/l_2\sin\alpha_1+\sin\alpha_2}{\cos^2\alpha_2} &-P\left(l_1/l_2\alpha_1+\alpha_2\right)  
	 \\[1mm]
	 Reut  &  P\sin\alpha_1 & P\alpha_1   & -Pl_1/l_2\sin\alpha_1  & -Pl_1/l_2\alpha_1\\[1mm] 
	\end{tabu}		
	\caption{\footnotesize The terms defining the external power in equation (\ref{nonlinearazzo}) and its linearized version (\ref{linearizzatissime}).}
	\label{externalpower}
\end{table}


The system of differential equations (\ref{linearizzatissime}) can be written in matrix form as 
\beq
\lb{gengei}
\bM 
\left[
\begin{array}{cc}
\ddot{\alpha}_1 \\
\ddot{\alpha}_2
\end{array}
\right] 
+
\left[
\begin{array}{cc}
c_1+c_2& -c_2\\
-c_2& c_2
\end{array}
\right] 
\left[
\begin{array}{cc}
\dot{\alpha}_1 \\
\dot{\alpha}_2
\end{array}
\right] 
+
\left(
\left[
\begin{array}{cc}
k_1+k_2& -k_2\\
-k_2& k_2
\end{array}
\right] 
+ P
\bL
\right)
\left[
\begin{array}{cc}
\alpha_1 \\
\alpha_2
\end{array}
\right] 
= 0
\eeq
For the considered four loads the load matrix $\bL$ is
\beq
\underbrace{
l_1\left[
\begin{array}{cc}
-1& 0\\
0& -l_2/l_1
\end{array}
\right] 
}_{Dead}
~~~~~~~
~~~~~~~
\underbrace{
l_1\left[
\begin{array}{cc}
-1& 1\\
0& 0
\end{array}
\right] 
}_{Ziegler}
~~~~~~~
~~~~~~~
\underbrace{
l_1\left[
\begin{array}{cc}
-1& 1\\
1& l_2/l_1
\end{array}
\right] 
}_{Contact}
~~~~~~~
~~~~~~~
\underbrace{
l_1\left[
\begin{array}{cc}
-1& 0\\
1& 0
\end{array}
\right] 
}_{Reut}
\eeq

Time-harmonic vibrations are analyzed near the equilibrium configuration, so that the Lagrangean parameters are now 
assumed to be harmonic functions of time
\beq
\lb{armonichetto}
\alpha_j = A_j \, e^{-i \Omega \,t},  ~~~j=1,2,
\eeq
where $A_j$ are (complex) amplitudes, $\Omega$ is the circular frequency, and $i$ is the imaginary unit ($i=\sqrt{-1}$), so that a substitution of eqn. 
(\ref{armonichetto}) into eqns. (\ref{linearizzatissime}) yields 
\beq
\lb{gengiazzo}
\left(-\Omega^2 \bM 
- i \Omega
\left[
\begin{array}{cc}
c_1+c_2& -c_2\\
-c_2& c_2
\end{array}
\right] 
+\left[
\begin{array}{cc}
k_1+k_2& -k_2\\
-k_2& k_2
\end{array}
\right] 
+ P
\bL 
\right)
\left[
\begin{array}{cc}
A_1 \\
A_2
\end{array}
\right] 
= 0 .
\eeq

Nontrivial solutions of the system (\ref{gengiazzo}) occur when the determinant of the matrix vanishes, a condition 
which is {\it identical for the cases of Ziegler and Reut}, but is different in the two other cases 
of frictionless contact and dead loading.

\subsection{Flutter and divergence instabilities in the perfectly elastic case for the Reut's column}

In the undamped case, $c_1=c_2=0$, the generalized eigenvale problem (\ref{gengiazzo}) can be rewritten as 
\beq
\label{geneig}
\left[\bK+P\bL - \Omega^2 \bM\right] \ba = 0,
\eeq
so that, since the mass matrix $\bM$ is real, symmetric and positive definite, its square root $\bM^{1/2}$ is invertible. Therefore, the  
generalized eigenvalue problem (\ref{geneig}) can be recast in a standard eigenvalue form 
\beq
\label{acazzo}
\left[\bM^{-1/2}\left(\bK+P\bL\right)\bM^{-1/2} - \Omega^2 \Id \right] \bM^{1/2}\ba = 0,
\eeq
showing that the only possibility of unsymmetry arises because of the nonsymmetry of $\bL$, 
possible only for the Reut or Ziegler loads. Therefore, only for these loads the squared circular frequency 
$\Omega^2$ can turn to be complex with non-null real part, which is the condition for
flutter instability. 

Due to the particular structure of the nonconservative loads so far considered, {\it the condition of vanishing for the
determinant of the matrix appearing in Eq. (\ref{acazzo}) is the same both for Ziegler and Reut systems.
Therefore, the conditions for flutter and divergence instability are the same.}

In the following, only the situation $l_1=l_2=l$, $k_1=k_2=k$, and $\rho_1 = \rho_2 = \rho$ is considered, because the experiments that will be reported in the following refer to this setting.
For the Reut system, the coefficient matrix becomes 
\beq
\lb{cavazza}
-\rho l^3 \Omega^2\left[
\begin{array}{ccc}
\ds \frac{4}{3} +  \frac{\rho_3l_3}{\rho l}   & \ds \frac{\rho_3l_3}{\rho l}+\frac{1}{2} \\ [5 mm]
\ds \frac{\rho_3l_3}{\rho l}+\frac{1}{2} & \ds \frac{\rho_3 l_3}{\rho l}+\frac{\rho_3 l_3^3}{\rho l^3 12}+\frac{1}{3}
\end{array}
\right]
+
k\left[
\begin{array}{ccc}
\ds 2 & -1 \\ [5 mm]
\ds -1 & 1
\end{array}
\right]
+P l 
\left[
\begin{array}{ccc}
\ds -1 & 0 \\ [5 mm]
\ds 1 & 0
\end{array}
\right] .
\eeq

The condition of vanishing of the determinant of the matrix (\ref{cavazza}) becomes
\beq
\lb{chi}
\Omega^4 (M_{11} M_{22}-M_{12}^2) + \Omega^2 \left[Pl(M_{22}+M_{12})-k(M_{11}+2M_{22}+2M_{12})\right] + k^2 =0,
\eeq
which is exactly the same expression that is obtained for the Ziegler load.

In particular, the flutter interval is defined by the condition that the discriminant of the equation (\ref{chi}) be null
\beq
P^2 - 2P\frac{k}{l} \frac{M_{11}+2M_{22}+2M_{12}}{M_{22}+M_{12}}  + 
\left(\frac{k}{l}\right)^2\left[\left(\frac{M_{11} +2M_{12}}{M_{22}+M_{12}}\right)^2 +4\right] = 0,
\eeq
leading to the two loads defining the range of flutter instability, which separates
the stable behaviour from the divergence instability
\beq
\lb{kaz}
P = \frac{k}{l} \frac{M_{11}+2M_{22}+2M_{12}
\pm 
2\sqrt{M_{11}M_{22}-M_{12}^2
}}
{M_{22}+M_{12}} ,
\eeq
where it should be noticed that the radicand is $\det \bM$, and thus always real and positive.

Eq. (\ref{kaz}) becomes for the considered matrix $\bM$ 
\beq
\lb{fluttazz}
P = \frac{k}{l} \frac{
36 +2\bar{\rho}\bar{l} (30+\bar{l}^2)\pm 
4\sqrt{7+3\bar{\rho}^2 \bar{l}^4+4\bar{\rho} \bar{l}(6+\bar{l}^2)} 
}{10+\bar{\rho}\bar{l}(24+\bar{l}^2)},
\eeq
where $\bar{l}=l_3/l_1$ and $\bar{\rho}=\rho_3/\rho$. 

In the special case in which the mass of the blade is negligible ($\bar{\rho}=0$), Eq. (\ref{fluttazz}) simplifies to
\beq
\lb{flot}
P = \frac{k}{l} \frac{
36 \pm 
4\sqrt{7} 
}{10} \approx \{2.542, 4.658\} \frac{k}{l} .
\eeq

\subsection{The destabilizing role of viscosity and the Ziegler paradox for the Reut system}

Due to the fact that {\it the determinant of the system (\ref{gengiazzo}) governing the linearized equilibrium is identical for both the Ziegler and Reut loadings}, 
it is expected that also in the Reut case the viscosity plays a destabilizing role even in the limit when the coefficients of viscosity 
are set to be equal to zero, the so-called \lq Ziegler paradox' 
\cite{ziegler_0, bolotin, bottema, kirillov_1, kirillov_libro, kirillov_2, kirillov_3, tommasini, luongo}, see the discussion reported in \cite{bigoni-cism}.

The hinges of the double pendulum are now assumed visco-elastic, so that 
the formulation (\ref{gengiazzo}) applies. 
Under the assumptions $l_1=l_2=l$, $k_1=k_2=k$ and now also $c_1=c_2=c$, 
and introducing the notation $\omega=-i\Omega$, the vanishing of the determinat of the 
coefficient matrix can be written as
\beq
p_0\omega^4 + p_1\omega^3 +p_2 \omega^2+p_3\omega+p_4=0,
\eeq
where
\beq
\begin{array}{cc}
	p_0=\det \bM,  ~~~~ p_1 = c\mu_1,  ~~~~ p_2 = -Pl\mu_2+k\mu_1+c^2,  ~~~~ p_3 = 2ck,   ~~~~  	p_4 =\ds  k^2, \\ [5 mm] 
	\mu_1= M_{11}+2M_{22}+2M_{12},  ~~~~~~~ \mu_2 = M_{22}+M_{12}.
\end{array}
\eeq

Following the Routh-Hurwitz criterion, stability occurs when the following three inequalities are all satisfied \cite{bigoni-cism, zieglerc} 
\beq
\lb{hur}
p_1>0, ~~~ p_1p_2-p_0p_3>0, ~~~(p_1p_2-p_0p_3)p_3-p_1^2p_4>0, ~~~ p_4>0.
\eeq

Condition (\ref{hur})$_3$ is the more restrictive and reads
\beq
P < \frac{1}{\mu_2} \left[ \frac{k}{l}\frac{\mu_1^2-4\det\bM}{2\mu_1}+ \frac{c^2}{l}\right],
\eeq
which becomes
\beq
\lb{pernac}
P < \frac{1}{10 +\bar{\rho}\bar{l}(24+\bar{l}^2)} \left[ \frac{k}{l}\frac{
296 +4\bar{\rho}\bar{l} (246+5\bar{l}^2)+\bar{\rho}^2\bar{l}^2(900+48\bar{l}^2+\bar{l}^4)
}{18+\bar{\rho}\bar{l}(30+\bar{l}^2)}+ \frac{12c^2}{\rho l^4}\right]. 
\eeq

In the special case when the mass of the blade is negligible ($\bar{\rho}=0$), Eq. (\ref{pernac}) simplifies to 
\beq
\lb{paradosso}
P <   \frac{
74 
}{45}\frac{k}{l}+ \frac{12\,c^2}{10 \rho l^4} \approx  1.644\frac{k}{l}+ 1.2\frac{c^2}{\rho l^4} ,
\eeq
an equation which clearly reveals the Ziegler paradox, because
the critical load in the limit of null viscosity, $c=0$, becomes smaller than 
the critical load for flutter evaluated assuming that the viscosity is \lq from the beginning' not present, Eq. (\ref{flot}).

\section{Reut's force from Coulomb friction}

In the proposed experimental set-up the Reut's force is obtained from frictional contact. The implementation of this force in the theoretical background  follows \cite{bigoninoselli}. 
When point $\bw$ is identified through the coordinate (\ref{oca}), the relative velocity at the blade/cylinder contact is 
\beq
\dot{\bw}_p = v_p \be_1+\dot{\bw.}
\eeq
The Coulomb rule for frictional contact  distinguishes between the two conditions when relative sliding does or does not occur, respectively 
\beq
\begin{array}{lll}
\dot{\bw}_p \scalp \be_1 \neq 0 & & \mbox{slip} \\ 
\dot{\bw}_p \scalp \be_1 = 0 & & \mbox{stick}
\end{array}
\eeq
so that the two frictional coefficients \lq static' $\mu_s$ and \lq dynamic', $\mu_d$ define the axial force $P$ from the vertical force $R$ transmitted orthogonally at the blade/cylinder contact as 
\beq
\lb{tzioo}
P = R \mu(\dot{\bw}_p \scalp \be_1) = R \left\{
\begin{array}{lll}
	\mu_d \,\,\, \mbox{sign}(\dot{\bw}_p \scalp \be_1) & & \mbox{slip} \\ 
\in [-\mu_s, \mu_s] & & \mbox{stick}
\end{array}
\right.
\eeq
where 
\beq
\lb{catzone}
\dot{\bw}_p \scalp \be_1  = v_p + l_1\sin\alpha_1 \left(-\dot{\alpha}_1+ \dot{\alpha}_2\right) . 
\eeq

\begin{figure}[h]
	\begin{center}		\includegraphics[width=0.9\textwidth]{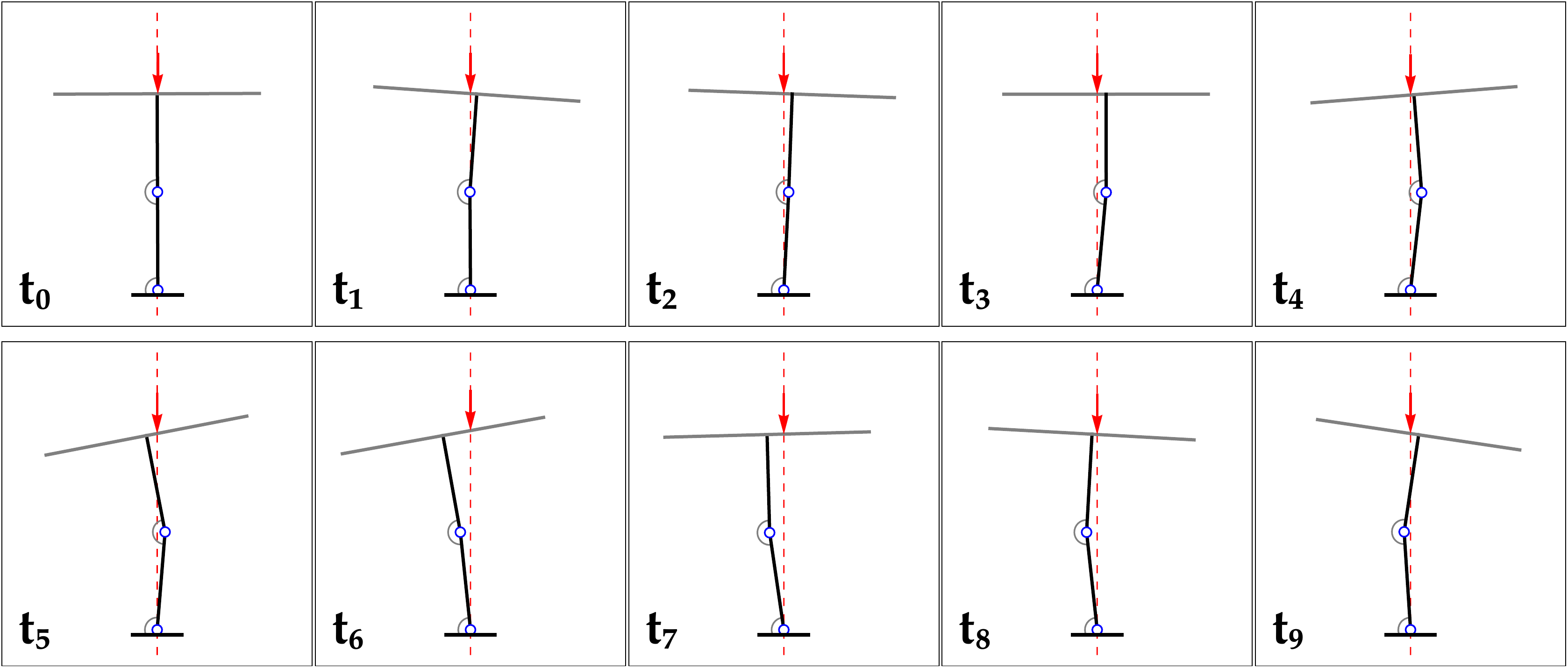}
		\caption{\footnotesize A sequence of images from a simulation of flutter instability in the Reut's column, obtained through numerical analysis of the nonlinear differential system (\ref{nonlinearazzo}) with initial conditions $\alpha_1=\alpha_2=-0.2$ and $\dot{\alpha}_1=\dot{\alpha}_2=0$. The load transmitted through friction in the straight configuration is 15 N 
		and the speed of the cylinder $v_p$ is equal to 20 mm/s.
	}
		\label{risultati_sim_flu}
	\end{center}
\end{figure}

\begin{figure}[h]
	\begin{center}		\includegraphics[width=0.9\textwidth]{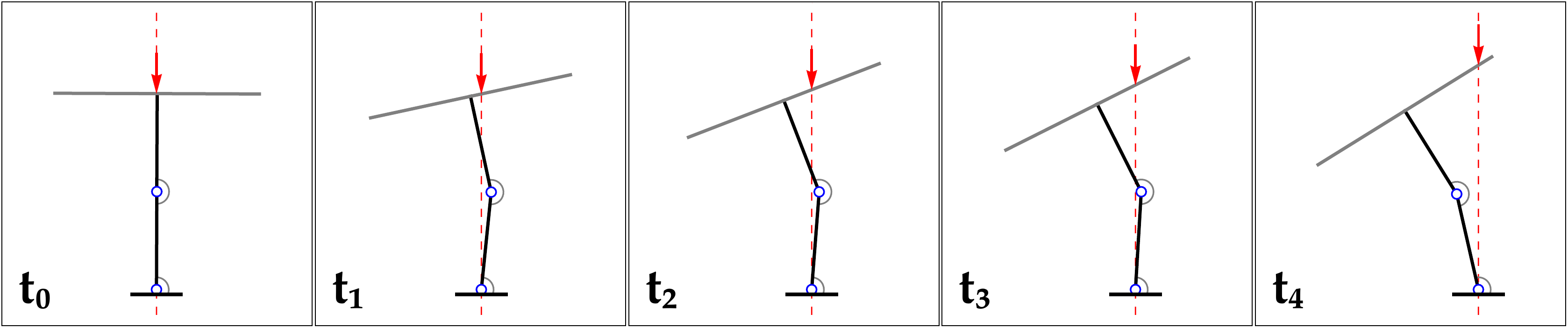}
		\caption{\footnotesize A sequence of images from a simulation of divergence instability in the Reut's column, obtained through numerical analysis of the nonlinear differential system (\ref{nonlinearazzo}) with initial conditions $\alpha_1=\alpha_2=0.2$ and $\dot{\alpha}_1=\dot{\alpha}_2=0$. The load transmitted through friction in the straight configuration is 60 N and the speed of the cylinder  $v_p$ is equal to 20 mm/s. 
		}
		\label{risultati_sim_div}
	\end{center}
\end{figure}

The numerical solution of the nonlinear differential system (\ref{nonlinearazzo}) with the force $P$ given by Eq. (\ref{tzioo}) was pursued with the 
same technique described for the Ziegler double pendulum \cite{bigoninoselli}, assuming for simplicity $\mu_s=\mu_d$ and introducing the regularization reported in  \cite{martinsa}, with the small parameter $\varepsilon$=0.04. 
Using for the geometrical and viscoelastic parameters the values representative of the experiments that will be presented in the next section, a sequence of images taken from a simulation of flutter instability and of divergence instability in the Reut's column is reported in Figs. \ref{risultati_sim_flu} and \ref{risultati_sim_div}, respectively.
The simulations were performed with initial conditions $\alpha_1=\alpha_2=\pm 0.2$ and $\dot{\alpha}_1=\dot{\alpha}_2=0$, at a speed $v_p$ of the cylindrical constraint equal to 20 mm/s. The load transmitted through friction in the straight configuration is 15 N and 60 N respectively for flutter and divergence.


The simulations show that flutter instability starts as an oscillation of increasing amplitude, but soon degenerates into a limit cycle motion, while 
divergence is characterized by a blowing-up dynamics, 
which is terminated when the force falls out of the blade.

\section{Experimental realization of the Reut's column}

The double pendulum with Reut's load was designed and realized in the \lq Instability Lab' of the University of Trento. Details on the 
construction are deferred to Appendix \ref{appendixb}. 
Difficulties arise in the practical implementation of the concept illustrated in Fig. \ref{lo_chiamavano_reut} from several sources. These were  already encountered in the realization of the Ziegler follower force, namely, the non-perfect adherence of friction to the Coulomb law and the non-null friction 
occurring in the mechanical system providing the compression to the head of the double pendulum. However, the most important difficulty is related to the fact that when the 
double pendulum is subject to a large deflection, the vertical transmission of the load becomes eccentric, so that a spurious torsion is induced. 
It is important to realize, however, that all the difficulties mainly concern the development of the instability, but not the onset of it. In fact, initially the 
double pendulum is in its straight configuration so that there are no eccentricities in load, and the measure of the critical loads for flutter and 
divergence is particularly accurate, as it is taken through a direct measure on the axis of the cylinder.

\begin{figure}[h]
	\begin{center}
		\includegraphics[width=0.9\textwidth]{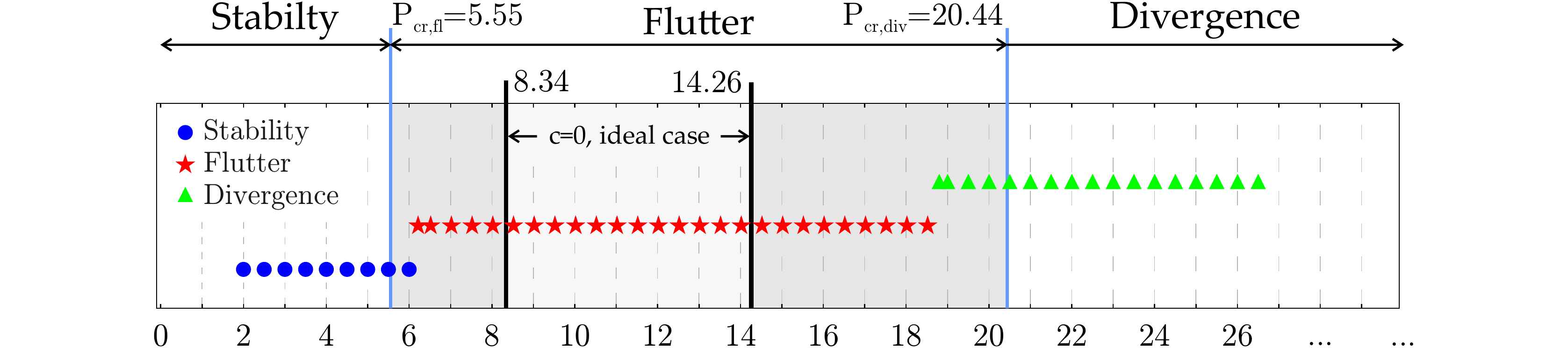}
		\caption{\footnotesize Comparison between the theoretical values and the experimental determinations of the threshold loads for flutter and for divergence. The experimental value of critical load for flutter (divergence), $P_{cr,fl}$=6.27 N ($P_{cr,div}$=18.72 N), lies between the two theoretical values referred to the viscous case, 5.55 N (20.44 N) and the purely elastic case, 8.34 N (14.26 N). Note that the experiments support the reduction in the critical load related to viscosity and thus the validity of the Ziegler paradox. 
		}
		\label{risultati1}
	\end{center}
\end{figure}

Experimental results for the determination of the critical loads for flutter and divergence instabilities are reported in Fig. \ref{risultati1}. 
Here the theoretical values for flutter instability depend on the account or not of the viscosity in the hinges. In the former case, called \lq ideal', 
the hinges are assumed purely elastic, while in the latter the hinges are viscoelastic.
The elasticity and viscosity of the hinges have been separately identified with {\it ad hoc} experiments (described in Appendix \ref{appendixb}), so that the 
values of the critical loads calculated with the viscoelastic model are believed to be tighter to the experiments. The theoretical values for the critical loads are
\beq
\lb{intervallaz}
\underbrace{\underbrace{5.55 \text{ N}}_{viscoelastic} ~~~~~ \underbrace{8.34 \text{ N}}_{elastic}}_{Flutter} ~~~--------~~~
\underbrace{\underbrace{14.26 \text{ N}}_{elastic} ~~~~~ \underbrace{20.44 \text{ N}}_{viscoelastic}}_{Divergence}
\eeq
which have been reported in Fig. \ref{risultati1} together with the measured values, reported as colored spots. 

The fact that the experimentally observed values of critical loads lie within the intervals (\ref{intervallaz}) provides 
the first experimental proof for the Reut's column of the detrimental effect of viscosity on the flutter load and, indirectly, of the validity of the Ziegler paradox for this structure. 

\begin{figure}[h]
	\begin{center}
		\includegraphics[width=0.8\textwidth]{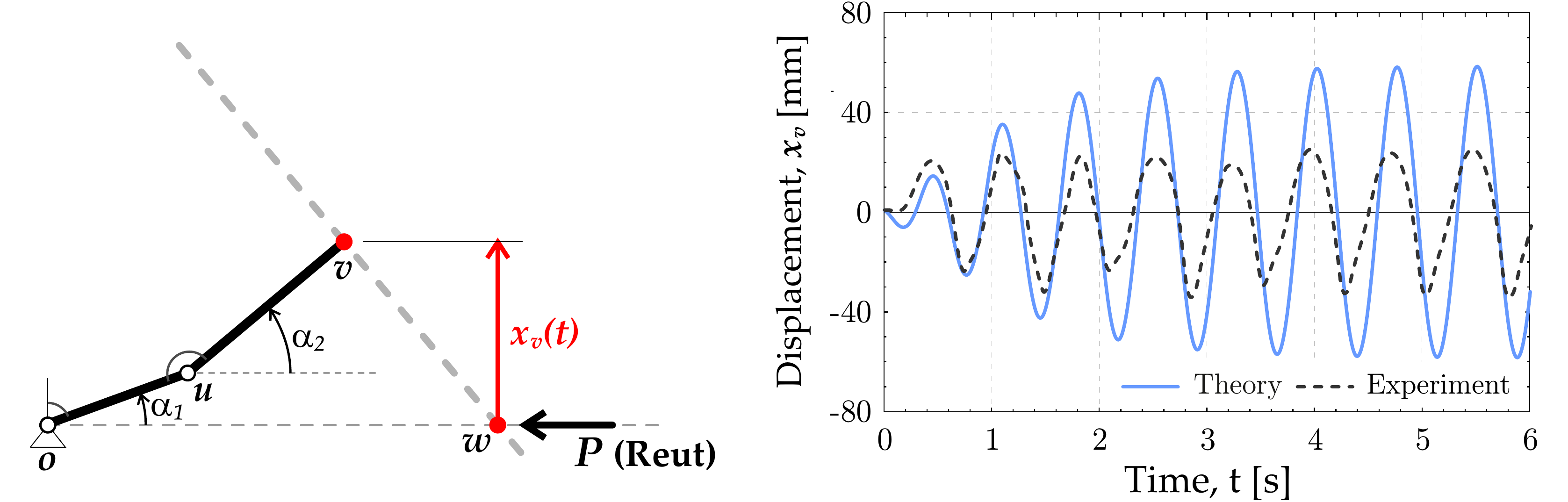}
		\caption{\footnotesize The oscillatory behaviour typical of flutter instability (at $P$=14 N), recorded as the transverse displacement component of the head of the double pendulum (speed of the cylinder $v_p$=20 mm/s). Note that a limit cycle is reached very soon, so that the structure behaves as a self-oscillating system.
		}
		\label{risultati2}
	\end{center}
\end{figure}

A record of the transverse displacement in time of the head of the double pendulum is provided in Fig. \ref{risultati2}, showing that the structure reaches a limit cycle almost immediately and thus behaves as a self-oscillating system \cite{jenkins}.

\begin{figure}[h]
	\begin{center}
		\includegraphics[width=0.8\textwidth]{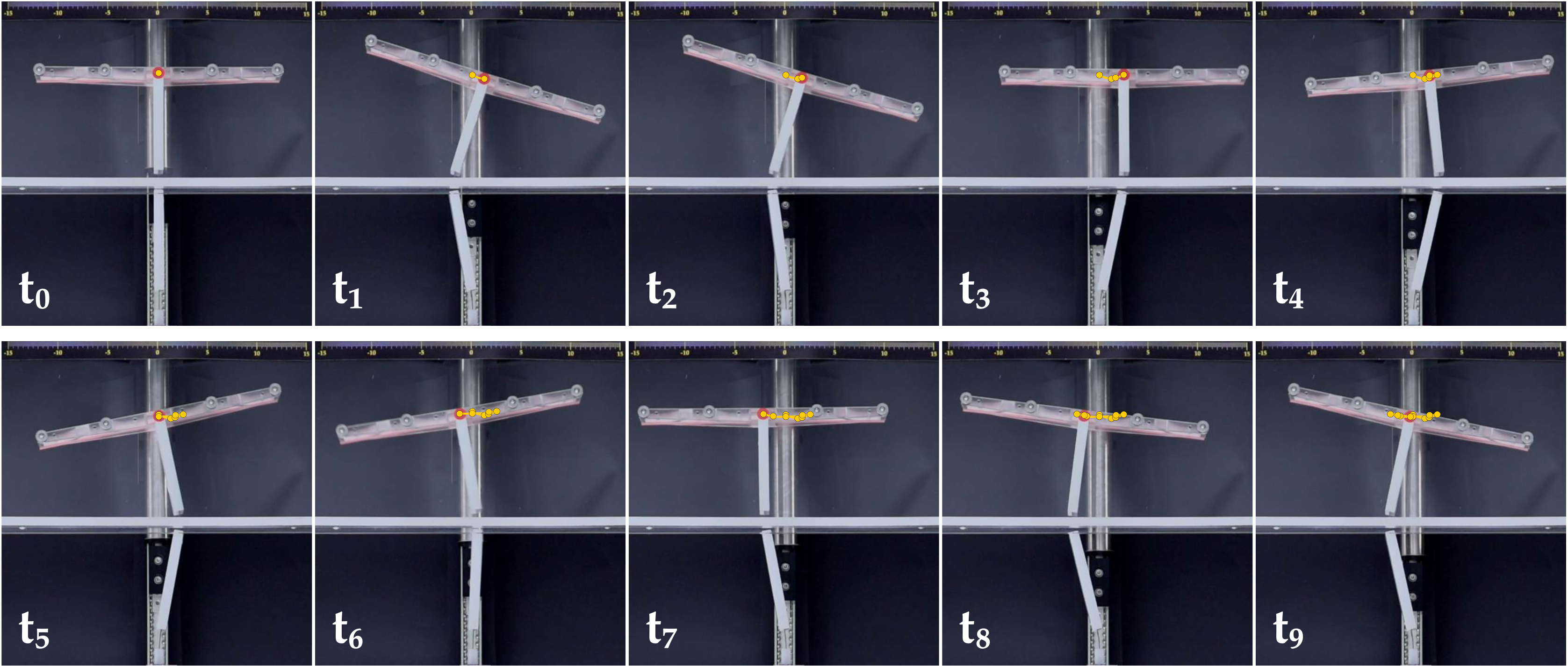}
		\caption{\footnotesize A sequence of photos documenting flutter instability in the Reut's column. The measured load in the straight configuration is $P=$14 N; a speed of $v_p$=20 mm/s 
			was imposed to the cylinder. Note the tracking of the head of the double pendulum (reported yellow).}	
	\label{uno}
	\end{center}
\end{figure}

\begin{figure}[h]
	\begin{center}
		\includegraphics[width=0.8\textwidth]{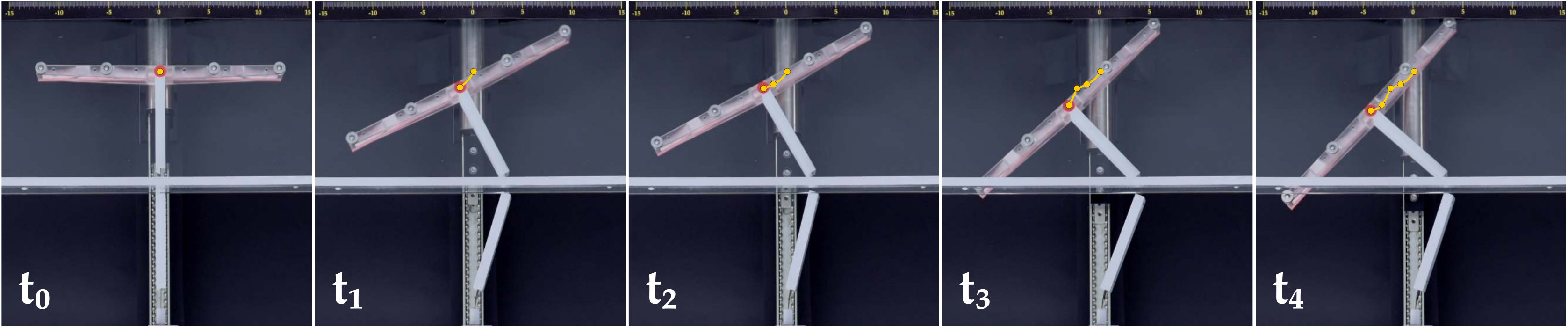}
		\caption{\footnotesize 
			A sequence of photos documenting divergence instability in the Reut's column. The measured load in the straight configuration is $P=$60 N; a speed of $v_p$=20 mm/s 
			was imposed to the cylinder. Note the tracking of the head of the double pendulum (reported yellow).
			}
		\label{due}
	\end{center}
\end{figure}

A sequence of photos documenting flutter instability and divergence instability are reported (in addition to Fig. \ref{sequenza}) respectively in Figs. \ref{uno} and \ref{due}, where also the tracking of the head of the double pendulum has been marked in yellow. 
Note that the experiments addressed to divergence instability were interrupted when the displacements became so large that the blade of the structure was 
escaping from the compressing glass (an event that would have ruined the testing machine). In this way it was not possible to document any oscillatory behaviour 
when divergence was observed. 


The results reported in Figs. \ref{risultati1} - \ref{due}, plus those collected in the movie of experiments included as supporting material (see also  http://ssmg.ing.unitn.it/), provide a 
coherent demonstration (i.) that the Reut's load can be realized, (ii.) that elastic structures subject to this load can suffer flutter and divergence instability, 
(iii.) that during flutter the structure behaves as a self-oscillating device and (iv.) that the viscosity decreases (increases) the threshold for flutter (for divergence).

\section{Conclusions}

The realization of the nonconservative load proposed near 80 years ago by Reut, namely, a force constrained to act along a straight line, was so far considered 
impossible. We have shown, theoretically and experimentally, how to produce this load on elastic structures, a finding which may be applicable to mechanical actuation, energy harvesting, and biomechanics.

\paragraph{Acknowledgments} D.B. acknowledges financial support from the PRIN 2015 ‘Multi-scale mechanical models for the design and optimization of micro-structured smart materials and metamaterials’ 2015LYYXA8-006. D.M. thanks support from the National Group of Mathematical Physics (GNFM-INdAM).
The authors also acknowledge support from the Italian Ministry of Education, University and Research(MIUR) in the frame of the \lq Departments of Excellence'  grant L. 232/2016. The authors thank Mr. Lorenzo P. Franchini and F. Vinante for assistance with the experiments.

IX 

\begin{appendices}
\section{The elastic double pendulum subject to four different loads: detailed analysis}\label{appendixa}

\subsection{Kinematics of the double pendulum}
The kinematics of the double pendulum is fully specified by the position, velocity and acceleration of points $\ba$, $\bb$, and $\bc$. For points 
$\ba$ and $\bb$ the following relations hold true
\beq
\lb{strasoccia2}
\begin{array}{lll}
	\ba  = a \be_r^{(1)} + \bo, \\ [3 mm]
	\dot{\ba} = a \dot{\alpha}_1 \be_\theta^{(1)} , \\ [3 mm]
	\ddot{\ba} = a \ddot{\alpha}_1 \be_\theta^{(1)} - a \dot{\alpha}_1^2 \be_r^{(1)},
	\\[1 mm] ~\\[1 mm]
	\bb  = l_1 \be_r^{(1)} + b \be_r^{(2)} + \bo, \\ [3 mm]
	\dot{\bb} = \ds l_1 \dot{\alpha}_1\be_\theta^{(1)} + b \dot{\alpha}_2 \be_\theta^{(2)}, \\ [3 mm]
	\ddot{\bb} = 
	l_1 \ddot{\alpha}_1\be_\theta^{(1)} - l_1 \dot{\alpha}_1^2\be_r^{(1)}
	+ b \ddot{\alpha}_2 \be_\theta^{(2)} -b \dot{\alpha}_2^2 \be_r^{(2)}, 
\end{array}
\eeq

The integrals defining the inertia in Eq. (\ref{virtuale}) can be solved to give
\beq
\begin{array}{lll}
	\ds \int_0^{l_1} \left(\ddot{\ba}\scalp\delta\ba \right) da &= \ds \frac{l_1^3}{3} \ddot{\alpha}_1\delta \alpha_1 \\ [5 mm]
	\ds \int_0^{l_2} \left(\ddot{\bb}\scalp\delta\bb \right) db & = \ds \delta\alpha_1 
	\left[
	l_1^2l_2\ddot{\alpha}_1+\frac{l_1l_2^2}{2}\ddot{\alpha}_2 \cos{(\alpha_1-\alpha_2)}
	+\frac{l_1l_2^2}{2}\dot{\alpha}_2^2\sin{(\alpha_1-\alpha_2)}
	\right] \\ [5 mm]
	& \ds +
	\delta \alpha_2 
	\left[
	\frac{l_1l_2^2}{2} \ddot{\alpha}_1 \cos{(\alpha_1-\alpha_2)} 
	- \frac{l_1l_2^2}{2} \dot{\alpha}_1^2\sin{(\alpha_1-\alpha_2)} 
	+\frac{l_2^3}{3} \ddot{\alpha}_2 
	\right],
	\\ [5 mm]
	\ds \int_{-l_3/2}^{l_3/2} \left(\ddot{\bc}\scalp\delta\bc \right) dc & = \ds 
	\delta\alpha_1 
	\left[
	l_1^2l_3\ddot{\alpha}_1+l_1l_2l_3\ddot{\alpha}_2\cos{(\alpha_1-\alpha_2)}
	+l_1l_2l_3\dot{\alpha}_2^2\sin{(\alpha_1-\alpha_2)}
	\right]  \\ [5 mm] 
	& \ds +
	\delta \alpha_2 
	\left[
	l_1l_2l_3 \ddot{\alpha}_1 \cos{(\alpha_1-\alpha_2)} 
	- l_1l_2l_3 \dot{\alpha}_1^2 \sin{(\alpha_1-\alpha_2)} 
	+\left(l_2^2l_3   +\frac{l_3^3}{12}   
	\right)\ddot{\alpha}_2 \right].
\end{array}
\eeq

\subsection{External power for three loads}

The power associated to external loads $\bP \scalp \delta \bw$ in Eq. (\ref{virtuale}) is now evaluated for the three cases of (i.) dead load, (ii.) Ziegler nonconservative load, and (iii.) frictionless contact with a guided weight. 

\subsubsection{Dead load (conservative)}

In the case of the dead load: (i.) the force $\bP$ is fixed at the point $\bv$ of the 
structure and remains (ii.) constant and (iii.) parallel to the $\be_1$-axis. 
Such a force follows the structure, because it is attached to the moving point $\bv$, 
and can be visualized as a weigth in a gravitational field (taking the $\be_1$-axis to be in the opposite direction of the field). 
The power of external load can be obtained from Eq. (\ref{kariko}) setting $c=0$ as
\beq
\bP\scalp\delta\bw = Pl_1  \sin\alpha_1\, \delta\alpha_1 +Pl_2 \sin \alpha_2 \delta\alpha_2.
\eeq

The dead load is conservative and in fact admits the potential
\beq
W(\alpha_1,\alpha_2) = P\left(l_1\cos\alpha_1+l_2\cos\alpha_2-l_1-l_2\right),
\eeq
so that 
\beq
\bP\scalp\delta\bw = -\frac{\partial W}{\partial \alpha_1} \delta\alpha_1
-\frac{\partial W}{\partial \alpha_2} \delta\alpha_2.
\eeq

The power of external load, linearized near the trivial equilibrium position, is for the dead load
\beq
\lb{kariko-dead}
\bP \scalp \delta\bw = Pl_1 \alpha_1\, \delta\alpha_1 +Pl_2 \alpha_2 \delta\alpha_2. 
\eeq

\subsubsection{Ziegler load (non conservative)}

In the case of the Ziegler load: (i.) the force $\bP$ is fixed at the point $\bv$ of the 
structure and remains (ii.) constant and (iii.) parallel to the bar $\bv-\bu$.
Such a force follows the structure and can be easily seen to be a nonconservative force
(Bigoni, 2018). 
The power of external load can be obtained from Eq. (\ref{kariko}) setting $c=0$ as
\beq
\bP\scalp\delta\bw = Pl_1 \sin(\alpha_1-\alpha_2) \, \delta\alpha_1,  
\eeq
an expression which cannot be derived from a potential.

The power of external load, linearized near the trivial equilibrium position, is for the Ziegler load
\beq
\lb{kariko-ziegler}
\bP \scalp \delta\bw = Pl_1 (\alpha_1-\alpha_2) \, \delta\alpha_1. 
\eeq

\subsubsection{Frictionless contact load with a dead weight (conservative)}

This load, which could be confused with the Reut load, can be imagined as a 
weight $P$ in a gravitational field acting on a constraint which 
forces the weight to slide along the $\be_1$-axis and stay in frictionless contact against the blade, namely, the bar orthogonal to $\bv-\bu$. Therefore, the latter bar is subject to a force orthogonal to it, free
of moving along the bar, and with horizontal component equal to $-P$. 

In the case of the frictionless contact: (i.) the force $\bP$ is free of sliding 
along the bar orthogonal to $\bv-\bu$, (ii.) remains orthogonal to it, while (iii.) 
$\bP\scalp\be_1 = -P$ is constant.
Such a force does not follow the structure, but has the same direction of the Ziegler load, 
although its modulus is not constant. In any case, the modulus of the Ziegler force and of the frictionless contact become identical in the linearized expression (\ref{linka}). 

The power of external load can be obtained from Eq. (\ref{kariko}), by setting for $c$ the value given by Eq. (\ref{oca}) and thus obtaining
\beq
\lb{pow}
\bP\scalp\delta\bw = Pl_1  \left(\sin\alpha_1-\cos\alpha_1\tan\alpha_2 \right) \delta\alpha_1 
-Pl_2 \left( 
\frac{l_1/l_2\sin\alpha_1+\sin\alpha_2}{\cos^2\alpha_2}
\right) \delta\alpha_2, 
\eeq
which is a conservative load, admitting the potential
\beq
W(\alpha_1,\alpha_2) = Pl_1\left(\cos\alpha_1+\sin\alpha_1\tan\alpha_2+\frac{l_2}{l_1\cos\alpha_2} - 1-\frac{l_2}{l_1}\right).
\eeq

The power of external load, linearized near the trivial equilibrium position, is for the frictionless contact load
\beq
\lb{kariko-contact}
\bP \scalp \delta\bw = Pl_1  \left(\alpha_1-\alpha_2 \right) \delta\alpha_1 
-P \left( l_1
\alpha_1+l_2 \alpha_2 
\right) \delta\alpha_2. 
\eeq

Note that there is another route to calculate the power of external load (\ref{pow}), which is to consider the power of the load $P$ sliding along the $\be_1$ axis. This movement can be calculated identifying 
$c$ with the value (\ref{oca}) before the differentiation in Eq. (\ref{strasoccia})$_1$ 
\beq
\lb{strasocm}
	\bc = l_1 \be_r^{(1)} + l_2 \be_r^{(2)} - 
	\left(
	\frac{l_1\sin\alpha_1+l_2\sin\alpha_2}{\cos\alpha_2}
	\right)
	\be_\theta^{(2)} + \bo,
	\eeq
so that 
\beq
\lb{tron}
\dot{\bc}= \ds l_1 \dot{\alpha}_1\be_\theta^{(1)} 
+ l_2 \dot{\alpha}_2\be_\theta^{(2)} 
+ \left(
\frac{l_1\sin\alpha_1+l_2\sin\alpha_2}{\cos\alpha_2}
\right) \dot{\alpha}_2 \be_r^{(2)}
+\dfrac{d}{d t}\left(
\frac{l_1\sin\alpha_1+l_2\sin\alpha_2}{\cos\alpha_2}
\right)  	
\be_\theta^{(2)}, 
\eeq
defines the velocity which produces the power directly from the load $P$. The scalar product of Eq. (\ref{tron}) with $-P\be_1$ provides the expression (\ref{pow}) for the power.

\subsection{Quasi-static bifurcations for the conservative systems}

A simple static analysis of the structure shown in Fig. \ref{reutinino} 
is sufficient to conclude 
that only the trivial (straight) configuration satisfies equilibrium for the cases of Ziegler and 
Reut loads, 
so that in these cases quasi-static bifurcations are excluded. 
The other two loads are different and non-trivial quasi-static bifurcations can be found, corresponding to quasi-static solutions for which the frequency vanishes, $\Omega=0$.

The loads for quasi-static bifurcations, occurring only for dead loading and frictionless contact loading, are independent of the viscosity of the hinges, because the matrix containing the coefficients $c_1$ and $c_2$ is multiplied by $\Omega$. 

The case of the dead load is well-know, while it can be remarkable 
that in the case of the frictionless contact bifurcations for both compressive and tensile loads may 
occur. 

When two bars of equal length are considered, $l_1=l_2=l$, together with identical spring stiffnesses, 
$k_1=k_2=k$, the dead load problem admits the following two compressive bifurcation 
loads
\beq
P_{dead}= \frac{k}{l} \frac{3\pm\sqrt{5}}{2}\approx \{0.3820, 2.618\} \frac{k}{l}, 
\eeq
while 
the frictionless contact problems admits the following tensile and compressive 
buckling loads (the former distinguished by the \lq$-$' sign)
\beq
P^{cr}_{contact}= \frac{k}{l} \frac{3\pm\sqrt{17}}{4} \approx \{-0.281, 1.781\} \frac{k}{l},
\eeq
which are both \lq critical', since they are both minimum values of loads which can be realized without adding further constraints to the system.



\section{The experimental set-up for the Reut's column}\label{appendixb}

\begin{figure}[h]
	\begin{center}
		\includegraphics[width=0.8\textwidth]{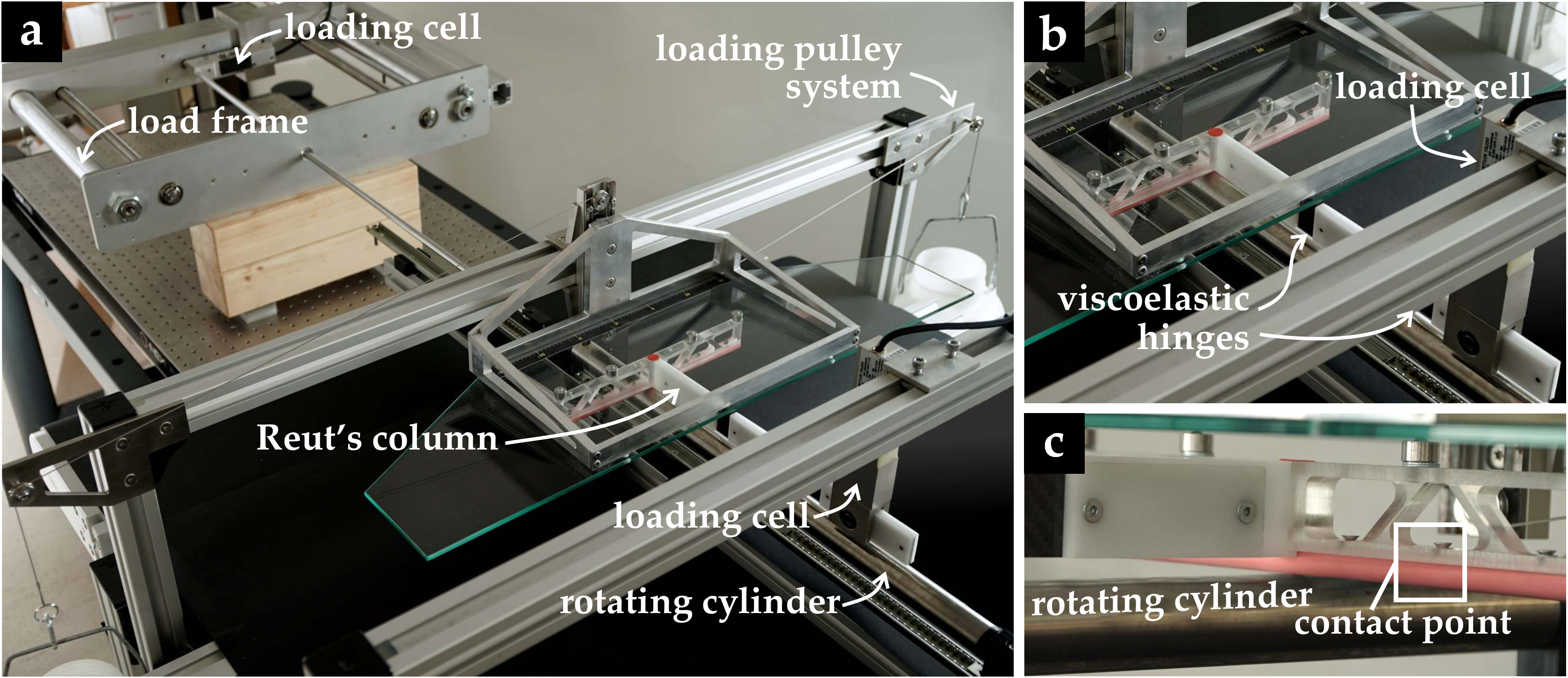}
		\caption{\footnotesize Experimental set-up showing the Reut's columns (detail b) in contact with the freely-rotating cylinder (detail c), moved against the structure at 
			fixed velocity by a testing machine (its load frame and loading cell  are visible, part a) turned horizontally. Note that the system used to load the head of the Reut's column is the same as that developed in \cite{bigkir1}.
		}
		\label{setup1}
	\end{center}
\end{figure}

Photos of the experimental set-up are reported in Fig. \ref{setup1}, where the \lq load frame' belongs to an electromechanical testing machine (Midi 10 from Messphysik Materials Testing) turned in a horizontal position. 

The elastic double pendulum was realized with two PMMA rigid bars (Young modulus $E$=2.35 GPa, $\rho_1$=$\rho_2$=0.322 kg/m; $l_1$=$l_2$=120 mm), while the transverse rigid bar  (dashed/grey in Fig. \ref{reutinino}) was 
realized as a truss structure in PMMA  ($\rho_3$=0.439 kg/m, $l_3$=250 mm), terminating with an aluminium bar (diameter $\phi$= 8mm) at its edge in contact with the cylinder.  The friction coefficient at the contact between the latter aluminium bar and the freely rotating cylinder was enhanced by wrapping the bar with a nitrile rubber NTR (thickness 1 mm, surface roughness 0.7 $\pm$ 0.1 $\mu$m, hardness 72 ShA, kindly provided by TyreF srl).

The Reut's load is transmitted to the double pendulum as a contact force generated through sliding against a stainless steel thin-walled cylinder (mounted on ball bearings, so that it is left free of rotating), moved at constant speed of $v_p$=20 mm/s by the above-mentioned electromechanical testing machine.  During the tests the reaction force at the fixed end of the pendulum and the force applied to the freely rotating cylinder 
were simultaneously acquired from two load cells (respectively a Mettler MT1041 RC 200N and a Mettler MT1041 RC 300N) with a  NI
CompactRio, interfaced with Labview
2018 (National Instruments). While conducting the experiments, high-speed movies were recorded (at 240 fps with a Sony PXW-FS5 high-speed camera) and employed to track the position in time of the head of the structure. 
Photos were taken with a Sony $\alpha$9 camera.

To transmit the frictional load, thin-walled tubes made of different materials (plastic, aluminium, galvanized steel and stainless steel) were used as the idealized cylindrical constraint. The most appropriate was found to be the stainless steel tube (diameter $\phi$=30 mm, thickness $t$=0.5 mm, length $L$=550 mm) which was placed on 
two linear guides (type Easy Rail SN22-80-500-610, from Rollon). Two ball bearings (model B626ZZ, from Misumi Europe)  were mounted at both ends of the tube to leave it free of rotating. 

As a system to vary the vertical force transmitted between the blade and the cylinder (so to vary the Reut's force via Coulumb friction) the same apparatus developed for the flutter machine was used (a detailed description is reported in \cite{bigkir1}). 
The friction coefficient measured at the blade/cylinder contact decreases with the vertical load within the 
interval $0.71 \pm 0.15$. This variation was taken into account in the simulations, but did not affect the measure of the critical loads for flutter and divergence, which was taken directly on the cylinder.

\begin{figure}[h]
	\begin{center}
		\includegraphics[width=0.8\textwidth]{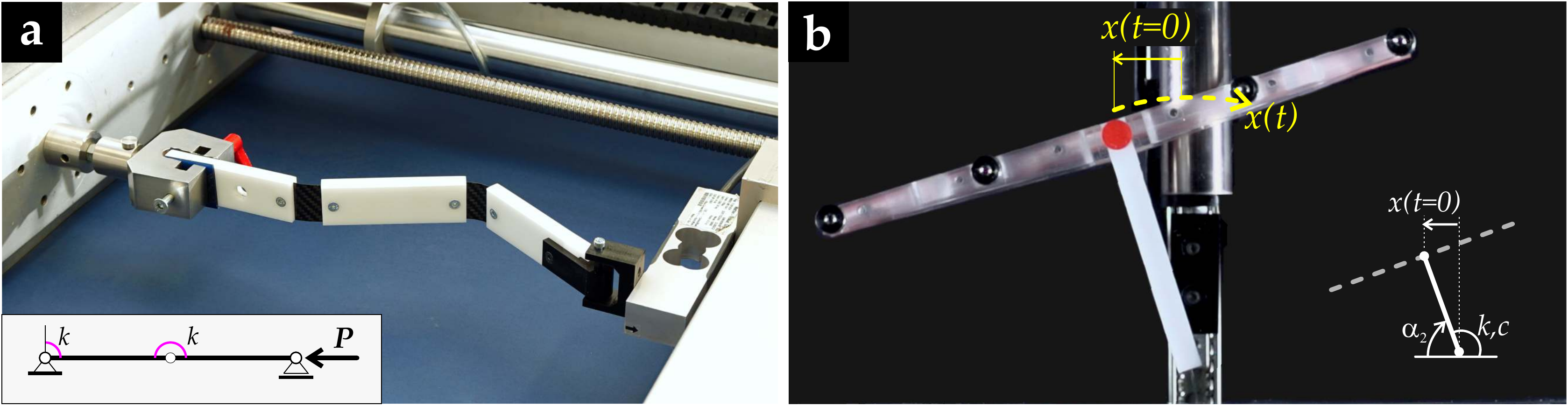}
		\caption{\footnotesize Identification of the hinges viscoelasticity through a buckling experiment to determine the stiffness $k_1=k_2$ (a) and through measurement of vibration damping to determine the viscous coefficients $c_1=c_2$ (b)}
		\label{fig_schema}
	\end{center}
\end{figure}

The viscoelastic hinges have been obtained with a strip of Carbon Fiber (Young modulus $E$=62 GPa) of thickness 0.41 mm, width 30 mm, and length $L$=20 mm. 
The elastic stiffness of the hinges was evaluated as $k_1$=$k_2$=$k$=0.531 Nm from a buckling test, Fig. \ref{fig_schema}a.  
The viscous coefficient of the hinges was identified to be $c_1$=$c_2$=$c$=0.012 Ns from 
a matching between simulated and measured free oscillations of the double pendulum with the first bar kept fixed (so that only rotation about the second hinge was allowed), Fig. \ref{fig_schema}b. 

\end{appendices}


\end{document}